\documentclass[11pt,a4paper]{article}

\usepackage{amsmath,amssymb}
\usepackage{epsfig,graphicx}
\usepackage{subfigure}
\usepackage{graphicx}
\usepackage{rotating}
\usepackage{cancel}
\usepackage{bm}
\usepackage{color}
\usepackage{comment}
\usepackage{braket}
\usepackage{cite}
\usepackage{psfrag}
\usepackage{slashed}
\usepackage{soul}
\usepackage{tikz}
\usepackage{hyperref}
\usepackage{bbold}
\usepackage[normalem]{ulem}

\renewcommand\[{\left[}

\newcommand{\exclude}[1]{}

\def\beq{\begin{equation}}
\def\eeq{\end{equation}}

\begin{document}
\title{
\Large{\textbf{On the time-dependent density of quadratically coupled dark matter around ordinary matter objects}}}

\author{Clare Burrage$^1$\footnote{\href{mailto:clare.burrage@nottingham.ac.uk}{clare.burrage@nottingham.ac.uk}}, Benjamin Elder$^2$\footnote{\href{mailto:b.elder@imperial.ac.uk}{b.elder@imperial.ac.uk}}, Yeray Garcia del Castillo$^{3,4}$\footnote{\href{y.garcia_del_castillo@unsw.edu.au}{y.garcia\_del\_castillo@unsw.edu.au}}, and Joerg Jaeckel$^{3}\footnote{\href{jjaeckel@thphys.uni-heidelberg.de}{jjaeckel@thphys.uni-heidelberg.de}}$\\[2ex]
\small{\em $^1$School of Physics and Astronomy, University of Nottingham,} \\
\small{\em University Park, NG7 2RD, Nottingham, United Kingdom}\\[0.5ex]
\small{\em $^2$Department of Physics, Imperial College London,} \\
\small{\em Prince Consort Road, SW7 2AZ, London, United Kingdom}\\[0.5ex]
\small{\em $^3$Institut f\"ur theoretische Physik, Universit\"at Heidelberg,} \\
\small{\em Philosophenweg 16, 69120 Heidelberg, Germany}\\
[0.5ex]
\small{\em $^4$School of Physics, The University of New South Wales,} \\
\small{\em Sydney NSW 2052, Australia}
\\[0.5ex]
}

\date{}
\maketitle

\begin{abstract}
\noindent
Wave-like dark matter may feature quadratic couplings to ordinary matter. This carries profound consequences for the phenomenologies of such models. It changes the dark matter density around dense objects made from ordinary matter such as planets and stars, thereby changing the sensitivity of direct detection experiments on Earth as well as implying forces on other ordinary matter objects in the vicinity. In this note we study the time dependence of the dark matter field around spherical objects of ordinary matter. This work indicates the time-scale on which accelerating objects settle into a stationary state and delineates the applicability of stationary solutions for experimental dark matter tests. We also use this to understand (and effectively eliminate) the infinities in energies, forces, and pressures that appear when naively comparing the total energy around objects with different size but the same total number of ordinary matter particles. 
\end{abstract}

\newpage

\tableofcontents
\newpage


\section{Introduction}\label{sec:intro}
Light and feebly interacting bosons provide for an interesting class of dark matter~\cite{Preskill:1982cy,Abbott:1982af,Dine:1982ah,Piazza:2010ye,Nelson:2011sf,Arias:2012az}  (see~\cite{Sikivie:2006ni,Jaeckel:2010ni,Marsh:2015xka,Hui:2021tkt,Adams:2022pbo,Antypas:2022asj} for some reviews). Due to their high occupation numbers they can be described as classical fields exhibiting a wave-like behavior~\cite{Sikivie:1983ip,Arias:2012az,Adams:2022pbo,Antypas:2022asj}. This allows for a wide variety of highly sensitive direct detection searches probing a variety of different couplings, see~\cite{Irastorza:2018dyq,Adams:2022pbo,Antypas:2022asj} for some examples and reviews.  

Many investigations focus on linear interactions with ordinary matter via different types of portal couplings (cf.~\cite{Okun:1982xi,Holdom:1985ag,Foot:1991kb,Binoth:1996au,Schabinger:2005ei,Patt:2006fw,Ahlers:2008qc,Batell:2009yf,Beacham:2019nyx}),  according to the scalar, pseudo-scalar or vector nature of the particle.
That said, for scalars and, more recently for pseudoscalars, quadratic interactions have also received some attention~\cite{Hees:2018fpg,Banerjee:2022sqg}.
They can be motivated as originating from a Higgs-portal~\cite{Binoth:1996au,Schabinger:2005ei,Patt:2006fw,Ahlers:2008qc,Batell:2009yf,Englert:2019eyl,Englert:2020gcp} but they also are a natural feature of axions and axion-like particles coupled to gluons~\cite{Hook:2017psm,Kim:2023pvt,Bauer:2023czj,Bauer:2024hfv,Bauer:2024yow} (albeit at higher order).

Quadratic interactions crucially differ from linear interactions in that objects of ordinary matter provide an effective potential for the dark matter particles. Depending on the sign of the interaction dark matter is then attracted or repelled from dense objects such as Earth or stars. This leads to a change in the local density of dark matter around compact objects such as planets and stars~\cite{Banerjee:2022sqg}. This crucially affects direct search experiments on Earth, but it also leads to phenomena such as effective dark matter density dependent forces between bodies of ordinary matter~\cite{Hees:2018fpg}.

Due to the oscillating nature of the dark matter field, there is an intrinsic time dependence to all these effects.\footnote{Of course this is also true in the case of linear couplings, where many detection methods are based on oscillating effects.} 
That said, if the ordinary matter object is at rest, or moving at a constant velocity with respect to the dark matter there often exist stationary solutions. These are the ones usually used when investigating quadratically coupled dark matter.  However, in most situations, e.g. the Earth rotating around the sun, this is only an approximation and it seems worthwhile to understand at which time-scale the stationary solutions are attained and become a good approximation. This is the first main question we want to ask in this note.

The second question arises from the apparently infinite energy difference between a situation with and without the object of ordinary matter present. Indeed this infinity seems to persist, even if we just change the volume of the ordinary matter object, while keeping its particle number fixed. To address this we study the energy and momentum flow in a finite volume.
From this we can see that the timescale over which the field relaxes to the stationary state increases with the size of the volume.  So the infinite energy difference is never encountered in practice, as it would take an infinitely long time.

The time dependence for a general situation is rather complicated. To keep things manageable we consider a simplified model wherein a sphere of homogeneous ordinary matter density appears, disappears, or changes its size. While this may not be fully realistic we nevertheless hope to, at least, capture the relevant time-scales.

In the next Sec.~\ref{sec:model} we make explicit our model conventions and review the stationary solutions from Ref.~\cite{Hees:2018fpg} featuring homogeneous boundary conditions at infinity. We then briefly discuss the Hamiltonian and illustrate the apparent problem of  infinite energies. In Sec.~\ref{sec:timedependence} we then describe the two methods by which we calculate the time-dependence of the scalar field. We then give results on the time dependence and an analytic estimate for the relevant time-scales. In Sec.~\ref{sec:enermom}, we explicitly consider the time evolution of the energy and momentum flows and use this to resolve the apparent infinities between solutions. In Sec.~\ref{sec:timeresults} we discuss the time-scales in relation to those relevant in experimental and observational settings before concluding in~\ref{sec:conclusions}. 

Note on conventions: we use the mostly minus metric signature, natural units $\hbar = c = 1$, and the reduced Planck mass $M_\mathrm{Pl} \equiv (8 \pi G)^{-1/2}$.

\section{Quadratically coupled scalar fields and their energy functional}
\label{sec:model}
We will assume that the dark matter particle is a light scalar field that is quadratically coupled to the matter density $\rho$,
\begin{equation}
    S = \int d^4x \left(  \frac{1}{2} (\partial \phi)^2 -\frac{1}{2} m^2 \phi^2 - \frac{\phi^2}{2 M^2} \rho \right)~.
    \label{action}
\end{equation}
Such a scenario can either be achieved by coupling the dark matter particle to the standard model or energy density explicitly (as shown here) or via a coupling to the Higgs boson, which {may be} integrated out at low energies to leave {an  effective coupling {similar to}} Eq.~\eqref{action}.

The resulting equation of motion is
\begin{equation}
    \left(\Box + m^2\right)\phi = -\frac{\phi}{M^2} \rho~.
    \label{EOM-scalar}
\end{equation}
With the choice of sign as given in Eq.~\eqref{action} the interaction between Standard Model matter and dark matter particles is repulsive. While we expect that most qualitative aspects of our discussion should also apply to the attractive\footnote{We note, however, that very interesting cases, such as the QCD axion or a gluon coupled axion-like particle are expected to feature an attractive effective interaction~\cite{Hook:2017psm}.} case, the discussion is more straightforward for the repulsive case as it avoids the existence of bound state solutions and instabilities that appear at larger values of the coupling in the attractive case.

\subsection{Analytical solutions for the stationary state}
To obtain an understanding of the behavior of the system it is instructive to consider the stationary solutions. First for the simple case of dark matter at rest without any ordinary matter present and then with a spherical object of ordinary matter of uniform density. The latter has been derived in Ref.~\cite{Hees:2018fpg} and used there to discuss the sensitivity of a variety of experiments. In Ref.~\cite{Banerjee:2022sqg} these solutions have been employed to consider changes in the sensitivity of dark matter detection experiments based on a linear coupling that are caused by the change in the dark matter density on the surface of Earth due to the quadratic coupling.

\subsubsection*{Homogeneous situation in absence of any matter}
The simplest case is for the unsourced equation, wherein $\rho = 0$.  The solution is straightforwardly seen to be
\begin{equation}
    \phi_\mathrm{homog}(t) = \phi_\mathrm{\infty} \cos( m t + \delta )~,
    \label{phi-solution-homogeneous}
\end{equation}
where $\phi_\mathrm{\infty}$ is a constant that sets the overall amplitude of the oscillations, and thereby the energy density of the dark matter field as $\rho_\mathrm{DM} = m^2 \phi_\infty^2$/2.

\subsubsection*{Stationary solution in presence of a spherical source object}
Let us now recall the solutions of~\cite{Hees:2018fpg} in presence of a spherical object of ordinary matter with density profile,
\begin{equation}
    \rho(\vec{x})=\rho\Theta(R-|\vec{x}|)~.
\end{equation}
We make an ansatz of the form
\begin{equation}
    \phi(\vec x, t) = \phi_\infty \cos(m t + \delta) \varphi(\vec x) ~.
    \label{eq:bgosc}
\end{equation}
The time derivatives cancel the $m^2$ term, and we are left with
\begin{equation}
\label{eq:eom}
    \vec \nabla^2 \varphi = \frac{\varphi}{M^2} \rho~.
\end{equation}
Outside the source, where $\rho = 0$, the solution is
\begin{equation}
    \varphi_\mathrm{out} = \frac{A}{r} + B~.
    \label{Poisson-portal}
\end{equation}
At $r = \infty$, we should have $\phi = \phi_\infty \cos (m t)$, so this sets $B = 1$.  The integration constant $A$ will be fixed later on by matching to the interior solution.

Now we solve inside the source, where $\rho = \mathrm{const}$.  The RHS looks like a mass term, so we have
\begin{equation}
    \varphi_\mathrm{in} = C\frac{e^{ \alpha m r}}{r} + D\frac{e^{- \alpha m r}}{r}~,
\end{equation}
where we have defined the dimensionless constant
\begin{equation}
    \alpha = \frac{\sqrt \rho}{m M}~.
    \label{alpha}
\end{equation}
Enforcing the boundary condition $\varphi'(0) = 0$ gives $C = -D$.  Finally, matching $\varphi, \varphi'$ at $r = R$, gives
\begin{equation}
    \phi_\mathrm{st}(r, t) = \phi_\infty \cos(m t + \delta ) \times \begin{cases}
        \frac{\mathrm{sech}(\alpha m R) \sinh(\alpha m r)}{(\alpha m r)}  & r \leq R~, \\ 
        1 - \frac{(\alpha m R) - \tanh(\alpha m R)}{(\alpha m r)} & r > R~.
    \end{cases}
    \label{phi-solution-quasistatic}
\end{equation}
This agrees exactly with the result for quadratic couplings in Ref.~\cite{Hees:2018fpg}.  We have written the spatially-dependent piece in terms of the dimensionless quantities $\alpha, (m r)$, and $(m R)$ for later convenience when we compare to numerical solutions.

We note that the deviation from the homogeneous background outside the source decays as $\sim 1/r$  and despite the field being massive there is no exponential suppression with the mass $m$ of the DM particle. This is because the presence of the source changes the energy and potential of real particles that already have an energy $m$ without the source.

\subsection{Energy of dark matter configurations}

The energy of a field configuration is, of course, given by the Hamiltonian corresponding to the action, Eq.~\eqref{action},
\begin{equation}
    H=\int d^{3}x\,\frac{1}{2}\left[\dot{\phi}^2+(\nabla\phi)^2+\left(m^2+\frac{\rho}{M^2}\right)\phi^2\right].
\end{equation}
For a stationary dark matter solution of the form of Eq.~\eqref{phi-solution-homogeneous} or Eq.~\eqref{eq:bgosc} with frequency $m$ this evaluates to
\begin{equation}
\label{eq:hamiltoniannonintegrated}
    H=m^2\frac{\phi_{\infty}^2}{2}\int d^{3}x\,  \varphi(\vec{x})^2
    +\frac{\phi^{2}_{\infty}}{2}\cos^{2}(mt+\delta)\int d^{3}x\,\left[(\nabla\varphi)^2+\frac{\rho}{M^2}\varphi^2(\vec{x})\right].
\end{equation}
Here, the first part is time-independent and the second part is non-vanishing only in the presence of a source.  For the homogeneous case, Eq.~\eqref{phi-solution-homogeneous}, the first term clearly gives the expected homogeneous energy density.

Looking at the second term of Eq.~\eqref{eq:hamiltoniannonintegrated} one may consider using partial integration and 
applying the equation of motion to further simplify it,
\begin{eqnarray}
\label{eq:hamiltonianpartial}
H&=&m^2\frac{\phi_{\infty}^2}{2}\int_{V} d^{3}x\,  \varphi(\vec{x})^2
    +\frac{\phi^{2}_{\infty}}{2}\cos^{2}(mt+\delta)\int_{V} d^{3}x\,\left[-\varphi(\Delta\varphi)+\frac{\rho}{M^2}\varphi^2(\vec{x})\right]
\\\nonumber    
&&\qquad\qquad\qquad\qquad\qquad\,
+\frac{\phi_{\infty}^2}{2}\cos^{2}(mt+{\delta})\int_{\partial V} d\vec{A} \varphi(\vec{x})\vec{\nabla}\varphi(\vec{x})~,  \\\nonumber
    &=& m^2\frac{\phi_{\infty}^2}{2}\int_{V} d^{3}x\,  \varphi(\vec{x})^2       +\frac{\phi_{\infty}^2}{2}\cos^{2}(mt+{\delta})\int_{\partial V} d\vec{A} \,\varphi(\vec{x})\vec{\nabla}\varphi(\vec{x})\,.
\end{eqnarray}
Here, we have made explicit the volume of integration and carefully included the boundary term arising in the partial integration. The simplification in the last line is due to the use of the equation of motion, Eq.~\eqref{eq:eom}.

We are now ready to evaluate the Hamiltonian for the stationary solutions given in Eq.~\eqref{phi-solution-quasistatic}. For simplicity we do so on a spherical volume centered on the matter distribution with a large radius $\Lambda\gg R$ and consider the leading terms in this radius,
\begin{align} \nonumber
    H =~ &m^2 \frac{\phi_\infty^2}{2} \left[ \frac{4\pi}{3} \Lambda^3 + \left(\frac{\tanh(\alpha m R)}{\alpha m R}-1\right) 4\pi R \Lambda^2+ {\mathcal{O}}(\Lambda) \right] \\
    &+ \frac{\phi_{\infty}^2}{2}\cos^{2}(mt+{\delta})\left[1-\frac{\tanh(\alpha m R)}{\alpha m R}\right] 4\pi R~.
    \label{eq:energytotal}
\end{align}
Crucially, the boundary term contributes the second line and is time-dependent. This implies that for a finite sphere energy is constantly moving in and out of a given volume.

\subsection{Infinite energy difference between stationary solutions}

The energy computed in the previous subsection grows with increasing volume of the sphere and indeed is infinite if we take the volume to infinity. This is not at all surprising as, asymptotically, we have a finite fixed dark matter density. This is reflected by the divergence with the volume, i.e. the first term on the right hand side of Eq.~\eqref{eq:energytotal}.

However, even subtracting this constant energy density the result Eq.~\eqref{eq:energytotal} still diverges with increasing radius $\Lambda$ of the considered sphere. Therefore the presence of a matter distribution changes the total energy in the stationary solution by an infinite amount. 
At first glance this may appear worrying, as we have not introduced any a priori divergent terms. However, it can be understood as an effect due to the infinite volume in the sense we can always move significant amounts of energy past any large but finite boundary surface.
For example, in the present case of a repulsive interaction between ordinary matter and dark matter, the leading order contribution to the energy difference is negative. Therefore if we ``appear'' a matter configuration we effectively have to expel an infinite amount of energy.
That said, this already hints at the relevance of time dependence to this problem. By causality, any effect that occurs at a distant boundary requires a significant time to take place.
This is why we investigate the time dependence of such a situation in the next section (Sec.~\ref{sec:timedependence}) and then explicitly discuss the energy and momentum flows in Sec.~\ref{sec:enermom}.

One may wonder whether appearing a matter configuration out of nothing is unphysical. However, the same problem of diverging energy differences persists, even if we only change the size of the matter configuration while keeping its total mass fixed. In this case the leading term quadratically diverging with the cutoff radius $\Lambda$ reads
\begin{equation}
    \frac{\partial H}{\partial R}=-4\pi \Lambda^2\tanh^{2}(\alpha m R)~.
\end{equation}
The infinity can again be resolved by considering that we are comparing two stationary solutions and taking into account that the transition between them takes time. As we will see later (cf. Appendix~\ref{app:contracting}) this ensures that quantities such as the pressure and energy flux remain finite.

In practice large objects are usually accelerating, for instance the Earth on its orbit around the Sun. Hence, it is not obvious that the solution for the dark matter field is well-captured by the stationary solution.\footnote{If the matter configuration moves with a constant velocity one can find suitable stationary solutions (see, e.g. Ref.~\cite{yeray}).} 
Aside from a better understanding of the apparent infinities this suggests that in practical situations we need to understand the relevant time-scales for the dark matter field's evolution when using stationary field configurations to calculate observable effects.

\section{Time-dependent solutions}\label{sec:timedependence}

As discussed in the previous section, between the two stationary solutions of a quadratically coupled scalar field an infinite amount of energy has to be moved. We are therefore interested in the time behavior of such systems.  Exact solutions to Eq.~\eqref{EOM-scalar} are very rare, so in this section we will construct approximate and numerical solutions.

For simplicity we consider a spherical source that appears at $t=0$ and then study the approach to the stationary solution, Eq.~\eqref{phi-solution-quasistatic}.
While certainly not fully realistic, this serves to give a rough understanding of the relevant time scales as well as the behavior of physical quantities such as the energy and momentum flow around the source object.
To put this into practice we will use two different techniques, a semi-analytic method based on techniques used in nonrelativistic quantum mechanics, as well as a fully numerical approach.

To be precise, we consider the case in which there is only the homogeneous solution, Eq.~\eqref{phi-solution-homogeneous}, at times $t < 0$, and then the source mass appears at $t = 0$.  We expect that the solution will tend towards $\phi_\mathrm{st}$ (Eq.~\eqref{phi-solution-quasistatic}) as $t \to \infty$.  We begin by discussing some particulars of the two solution methods, comparing the results, and finally we summarize main observed behaviors at the end of the section.

\subsection{Semi-analytical approach}
\label{subsec:semi-analytical}
In order to study non-stationary solutions of our system (solutions with a non-oscillating time dependence), we can use a semi-analytical method based on calculating the eigenfunctions of the differential Eq.~\eqref{EOM-scalar}, for a spherical source with density $\rho$ and radius $R$. This is both physically and mathematically equivalent to the quantum mechanical problem of finding the energy eigenstates for a finite spherical potential well.
The resulting set of eigenfunctions, $\Tilde{\phi}_{k }$, 
can be found in Appendix~\ref{eigenfunctions}. It is a complete basis in the space of functions, and any initial field configuration $\phi_0 (\vec{r},0)$ can be projected onto it. Since the time evolution of the eigenfunctions is known, the time evolution for the initial configuration can be straightforwardly computed. The general result then reads
 \begin{equation}
\phi_0(\vec{r},t)=\int \cos(\omega(k) t)\Tilde{\phi}_{k }(\vec{r})\braket{\Tilde{\phi}_k|\phi_0} \,dk^3~,
\label{418}
 \end{equation}
where
 \begin{equation}
\braket{\Tilde{\phi}_k|\phi_0}=\int\overline {\Tilde{\phi}_k(\vec{r})}\phi_0 (\vec{r},0)\,dV~,
\label{419}
 \end{equation}
 and $\omega(k)$ is the frequency associated with the different eigenfunctions.

In the concrete case of an appearing source at $t=0$, the projection in Eq.~(\ref{419}) diverges since the initial field configuration, given by Eq.~(\ref{phi-solution-homogeneous}), is a non-normalizable function. However, a cut-off can be introduced  to normalize the function $\phi_0(\Vec{r},0)=e^{-\epsilon r}\phi_\mathrm{homog}(0)$, making the projection finite. For an appropriate choice, the cut-off will not affect the results. In practice, as long as $\frac{1}{\epsilon}\gg r$ and $\frac{1} {\epsilon}\gg t$ the approximation is good, because the information from the points far away where $r\approx \frac{1}{\epsilon}$ cannot reach the spatial region in which we are interested.

Implementing this method is quite simple and does not require too much computational effort. For the homogeneous initial field configuration we are considering, the projection in Eq.~(\ref{419}) can be done completely analytically. On the other hand, the integral in Eq.~\eqref{418} needs to be evaluated numerically. Nevertheless this integral reduces to a one-dimensional one due to the spherical symmetry of the problem. Using this the integral can be computed using an adaptive quadrature method to yield the time evolution of the field. 

Equation \eqref{418} is a weighted sum over the continuous set of eigenfunctions, where $k$ can be any positive real number. Since the projection, after introducing the cut-off, is a well-behaved function as $k$ goes to infinity, the numerical integration can be done by taking a reasonably high value for the upper integration limit. In the adaptive quadrature method, this upper limit is set by the requirement of a given relative tolerance for the integral, which we have chosen to be $10^{-6}$.

\begin{figure}
    \centering
    \begin{tikzpicture}
\node at (0, 4.75*2){
\includegraphics[width=0.32\linewidth]  {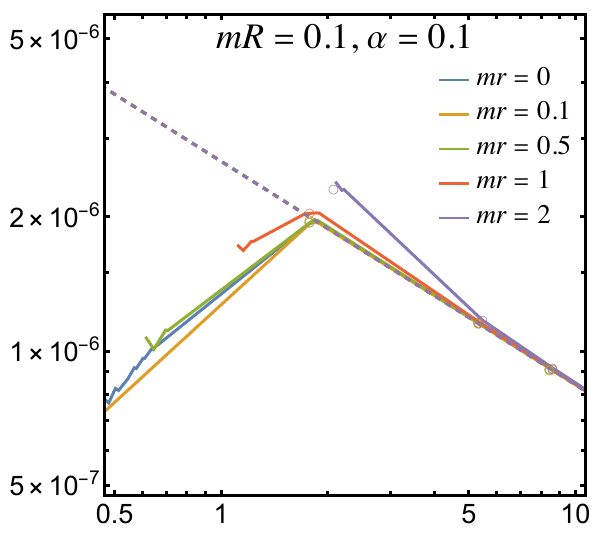}
\includegraphics[width=0.32\linewidth]  {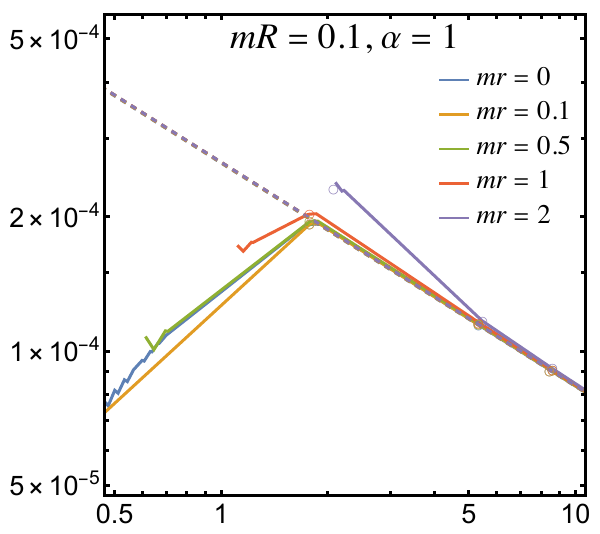}
\includegraphics[width=0.32\linewidth]  {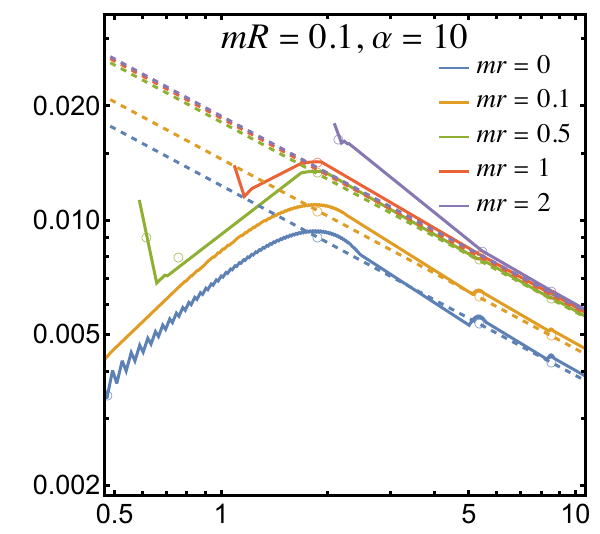}};

\node at (0, 4.75){
\includegraphics[width=0.32\linewidth]{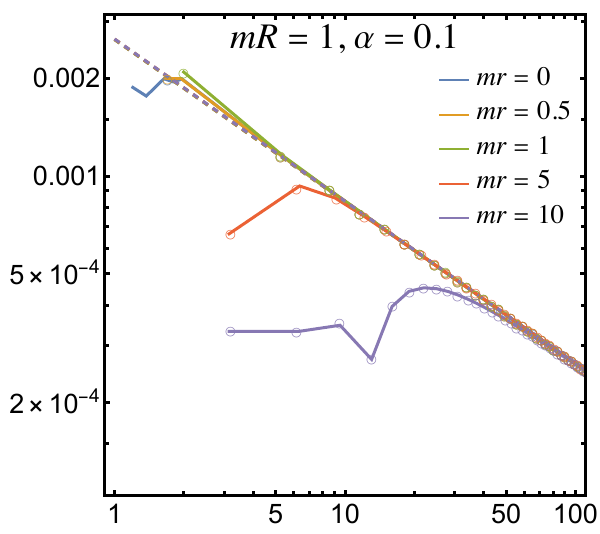}
    \includegraphics[width=0.32\linewidth]{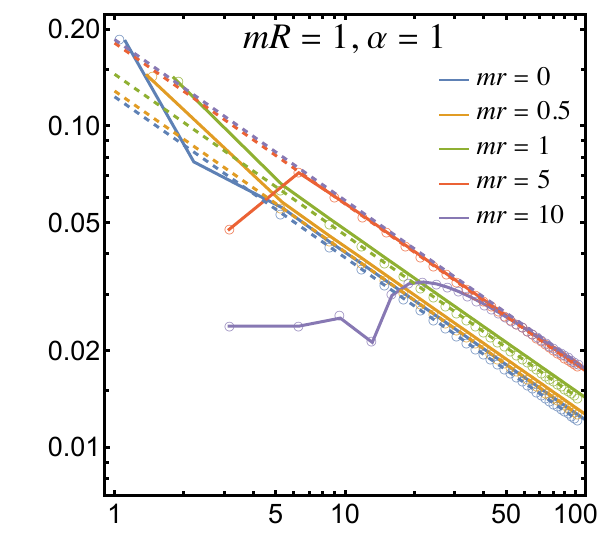}
    \includegraphics[width=0.32\linewidth]{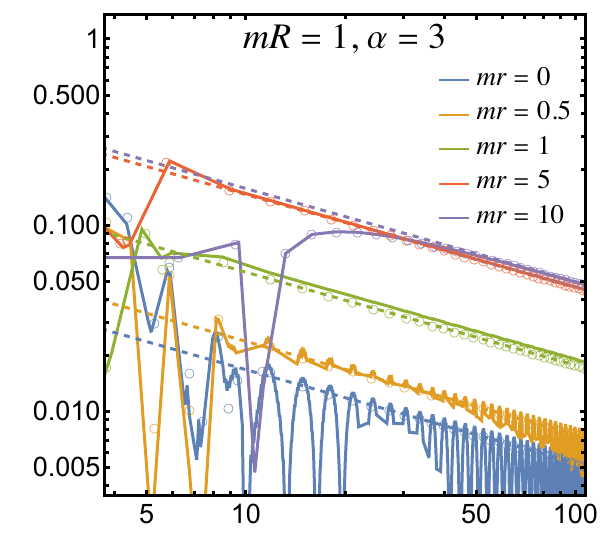}};

\node at (0,0){
\includegraphics[width=0.32\linewidth]{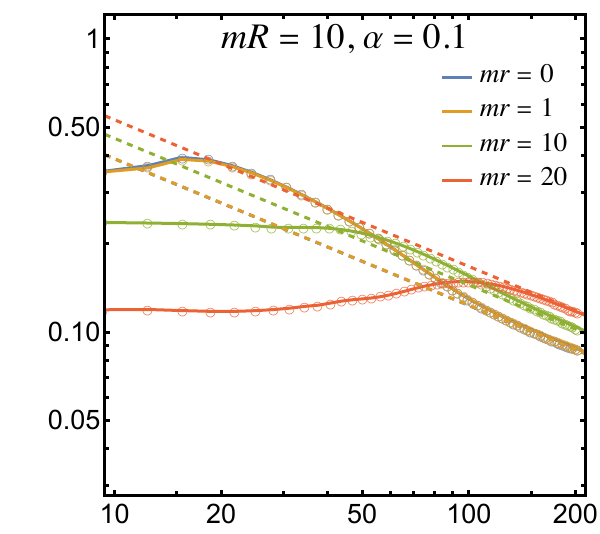}
    \includegraphics[width=0.32\linewidth]{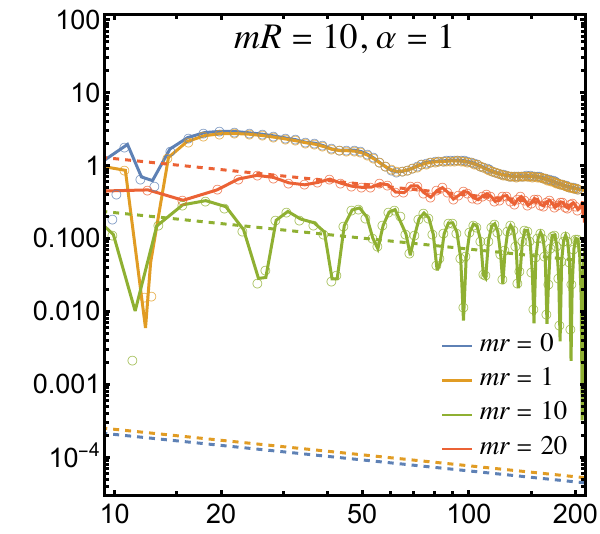}
    \includegraphics[width=0.32\linewidth]{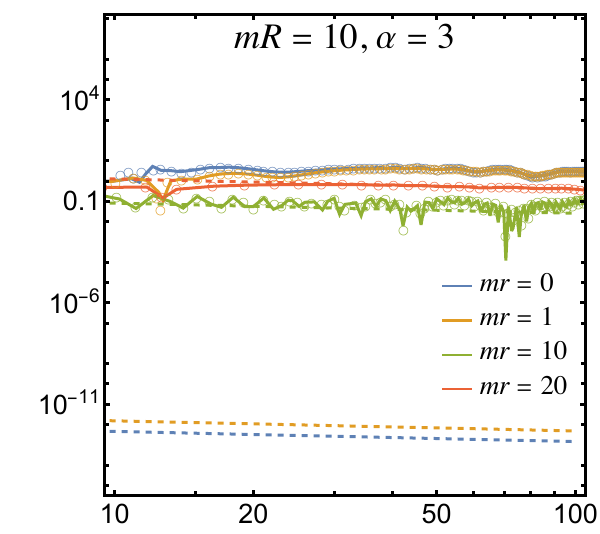}};

\node at (-4.75,-2.5){Time $m t$};
\node at (0.5,-2.5){Time $m t$};
\node at (5.75,-2.5){Time $m t$};

\node[rotate=90] at (-7.75, 0){Osc. amplitude of $\varphi$};
\node[rotate=90] at (-7.9, 4.75){Osc. amplitude of $\varphi$};
\node[rotate=90] at (-7.9, 4.75*2){Osc. amplitude of $\varphi$};
    
\end{tikzpicture}
\caption{\small Comparison between numerical (colored solid curves) and semi-analytic (colored circles) solutions, for a fixed source radius and a range of source densities $0.1 < \alpha \leq 3$.  The vertical axis is the amplitude of the oscillations in $\varphi \equiv \phi - \phi_\mathrm{st}$.  The colored dashed lines show the asymptotic solution given by Eq.~\eqref{sacada} and shows that the oscillations damp away as $(m t)^{-1/2}$. The numerical and semi-analytical approach agree quite well for the parameters represented.  Excellent agreement is found with the asymptotic solution, except deep inside large and dense sources, where  the field takes much longer to react to the appearance of the source. }

    \label{fig:amplitude-comparison}
\end{figure}

As noted earlier, we expect that the field will tend towards the stationary configuration $\phi_\mathrm{st}$, Eq.~\eqref{phi-solution-quasistatic}, at late times. Our central observation is to quantify how the solution approaches the stationary solution. This is shown in Fig.~\ref{fig:amplitude-comparison}. Using the black dashed line to guide the eye we can see that the field tends to the quasistatic configuration,  scaling as $\sim (m t)^{-1/2}$ at late times. 

This behavior can also be inferred using Eq.~\eqref{418} to obtain an analytical approximation for the long-time behavior. Expanding Eq.~\eqref{418} around $k=0$, the expression simplifies and an analytical approximation may be obtained. This expansion is justified because at late times the eigenfunctions close to $k=0$ are the ones that dominate the behavior of the field. 
The full derivation of the analytical formula can be found in the Appendix~\ref{app:asymptotic}. Here, we only show the final expression:
\begin{equation}
\phi_0(\vec{r},t) \approx  \phi_\mathrm{st}(r, t) \left(1+ \frac{\alpha m R-\tanh(\alpha m R)}{\alpha} \sqrt{\frac{2}{\pi m t}}\right)~.
\label{sacada}
\end{equation}
This analytical expression describes the asymptotic behavior of the field, that is, in the limit $t \to \infty$ the field tends to the quasi-static configuration as it has been already indicated. Note also that the spatial dependence of the scaling is now explicit, and is given by the amplitude of the quasi-static solution. Finally, the time dependence obtained explicitly demonstrates the $(m t)^{-1/2}$ scaling that was observed in Fig.~\ref{fig:amplitude-comparison}. We will see shortly that this expression agrees well with fully numerical solutions. Later in Sec.~\ref{sec:timeresults}, we will use this result to estimate the relevant time scales for a realistic situation.

\subsection{Numerical Approach}\label{subsec:numerical}
To validate the semi-analytical approach we have also performed fully numerical calculations of the time-dependence.

To solve Eq.~\eqref{EOM-scalar} numerically, it is helpful to redefine our variables in the following way:
\begin{equation}
    \hat \phi = \phi / \phi_\infty~, \quad \hat x^\mu = m x^\mu~.
    \label{phi-scaling}
\end{equation}
The advantage of this is that all terms in the equation of motion are now dimensionless:
\begin{equation}
    (-\hat \Box - 1) \hat \phi = \alpha^2 \hat \phi~.
    \label{EOM-numerical}
\end{equation}
We have eliminated explicit dependence on $m$ and $\phi_\infty$, removing the need to scan over these parameters.  Put another way, given a solution $\hat \phi$ we can use it to construct entire families of solutions using the scaling relations of Eq.~\eqref{phi-scaling}.

We solve this equation for spherically symmetric configurations, so
\begin{equation}
    - \hat \Box \hat \phi = -\frac{d^2}{d \hat t^2} \hat \phi + \frac{1}{\hat r^2} \frac{d}{d \hat r} \left(\hat r^2 \frac{d \hat \phi}{d \hat r} \right)~.
\end{equation}
This PDE is solved on a finite domain $0 < \hat r < \hat r_\mathrm{max}$.  Furthermore, it will be subjected to the initial condition $\hat \phi(\hat t = 0, \hat r) = \hat \phi_\mathrm{IC}(\hat r), \dot{\hat \phi}(\hat t = 0, \hat r) = \dot{\hat \phi}_\mathrm{IC}(\hat r)$, and the boundary conditions
\begin{align} \nonumber
\frac{d}{d \hat r}\hat \phi(\hat t, 0) &= 0~,\\
\hat \phi(\hat t, \hat r_\mathrm{max}) &= \hat \phi_\mathrm{BC}~,
\label{boundary-conditions-numerical}
\end{align}
where we choose $\hat{\phi}_{\mathrm{BC}}$ to be the appropriately normalized version of Eq.~\eqref{phi-solution-homogeneous}.
Likewise, the initial condition $\hat \phi_\mathrm{IC}$ is obtained from the homogenous solution Eq.~\eqref{phi-solution-homogeneous} at $t = 0$.

To solve this numerically we employ the ``method of lines'', which works as follows.  We discretize space into $N$ segments of width $\hat{dr} = \hat r_\mathrm{max} / N$, allowing us replace the spatial derivatives with second-order finite difference operators.  This converts a single PDE into $N-2$ coupled second-order ODEs, with the two remaining equations provided by the boundary conditions.  These ODEs are integrated with a standard ODE solver.

The coordinate singularity at $r = 0$ is dealt with neatly via the boundary condition in Eq.~\eqref{boundary-conditions-numerical}.  Once discretised, this boundary condition gives
\begin{equation}
    \hat \phi(\hat t, 0) = \hat \phi(\hat t, \hat{dr})~.
\end{equation}
Thus the coordinate singularity at the origin is never encountered.

The boundary condition at $\hat r_\mathrm{max}$ requires some care, as we wish to avoid introducing numerical artifacts originating from the finite size of the volume.  Fortunately, we are mainly interested in the transient behavior of the solution, so it is not necessary to simulate the system arbitrarily far into the future.  In our units the sound speed is 1, so any artifacts from the boundary condition will take at least $\hat t = \hat r_\mathrm{max}$ to propagate to the origin.  For a given location $\hat r$ we sidestep the issue completely by restricting our attention to times $\hat t < \hat r_\mathrm{max} - \hat r$.

Setting $\hat dr$ also requires some care.  We note that the source term in Eq.~\eqref{EOM-numerical} appears like a mass, so the effective Compton wavelength of the scalar field inside the source is $\lambda_\mathrm{Compton} = (1 + \alpha^2)^{-1/2}$.  This is the characteristic length scale for spatial variations inside the source, so we should use a resolution that is sufficiently smaller than this.  In practice we choose $\hat{dr} \lesssim \lambda_\mathrm{Compton} / 10$.  We also require sufficient resolution to capture the field variations within the source mass, regardless of the field's Compton wavelength, so we set
\begin{equation}
    \hat{dr} = \frac{1}{10}\min\left(\lambda_\mathrm{Compton}, R \right).
\end{equation}
The algorithm runs in a reasonable amount of time on a personal computer so long as the number of grid lines is $N \lesssim 10^3$.

\bigskip

As a check of this numerical strategy, we first consider a related but slightly simpler problem: that of a disappearing source.  That is, we begin with the solution for an extended body in an oscillating background field, and then at $t = 0$ the source mass vanishes.  We then study how the field relaxes to the homogeneous background oscillation of Eq.~\eqref{phi-solution-homogeneous}.  In this case it is possible to develop an approximate solution for the field at the origin, $\phi(r = 0, t)$ for small $mt$.  This is derived in Appendix~\ref{app:disappearing-source}, along with a description of the numerical setup and results for the problem.  We find excellent agreement between the numerical and analytical solutions.  Furthermore we find a similar $\sim(m t)^{-1/2}$ long-time behavior as already discussed in Sec.~\ref{subsec:semi-analytical} for the case of an appearing source.

\bigskip

Let us now turn to our main problem of interest, the appearing source.  Due to the linearity of the equation of motion it is useful to consider the difference to the stationary solution. To do so we define our field variable as, 
\begin{equation}
    \hat \phi = \hat \phi_\mathrm{st} + \varphi~.
\end{equation}
With this definition we expect the amplitude of $\varphi$ to decay in time as the field approaches the stationary solution. 
The equation of motion for $\varphi$ then reads,
\begin{equation}
    (- \hat \Box - 1) \varphi = \hat \rho \varphi~,
\end{equation}
where $\hat \rho = \alpha^2 \Theta(\hat R - \hat r)$, and $\Theta(x)$ is the Heaviside step function~.
Our initial condition for $\varphi$ is given by
\begin{equation}
    \varphi(t \leq 0) = \hat \phi_\mathrm{homog} - \hat \phi_\mathrm{st}~.
\end{equation}
The boundary conditions are $\partial_r \varphi(t, r = 0) = 0$ and  $\varphi(t, r_\mathrm{max}) = 0$.

\begin{figure}
    \centering
    \includegraphics[width=0.8\linewidth]{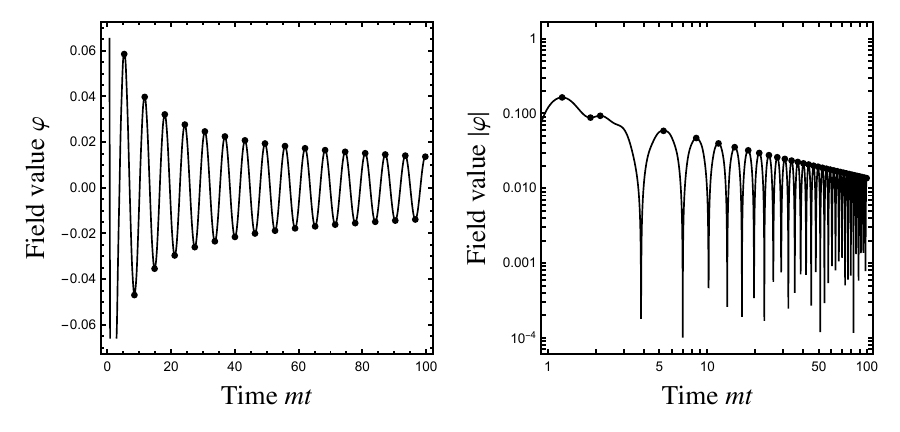}
    \caption{\small Field solutions for $mR = \alpha = 1$, at the location $r = 0$. Extremal points are indicated with solid points.  The same solution is shown in both plots, on linear axes (left) and logarithmic axes (right).  The extremal points on logarithmic axes make it clear that the amplitude of the oscillations in $\varphi$ damp away as a power law.} 
    \label{fig:amplitude-explanation}
\end{figure}

A sample solution is plotted in Fig.~\ref{fig:amplitude-explanation}.  The numerical solution is shown as a solid line, on linear axes (left) and logarithmic axes (right).  Our primary interest is not in the oscillations, but rather the rate at which the oscillations damp away.  To extract this information from the numerical solutions, we detect local extrema which are indicated with solid points.  Putting the absolute value of these points on a logarithmic scale allows one to clearly see the power law behavior of this damping.  In the plots of our main results, we will omit the oscillatory solutions and only show the decay of the oscillations' amplitude over time.

\begin{figure}
    \centering
    \includegraphics[width=0.4\linewidth]{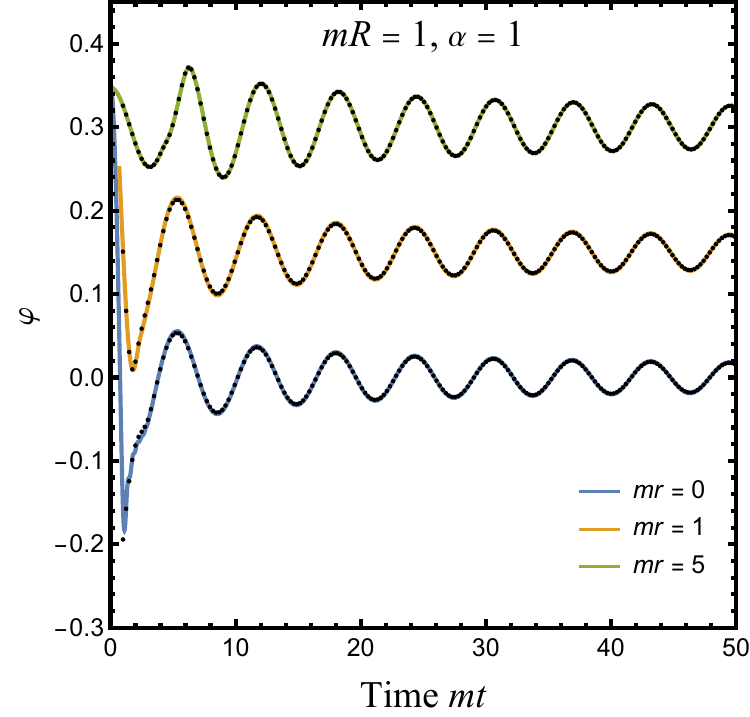}
    \includegraphics[width=0.4\linewidth]{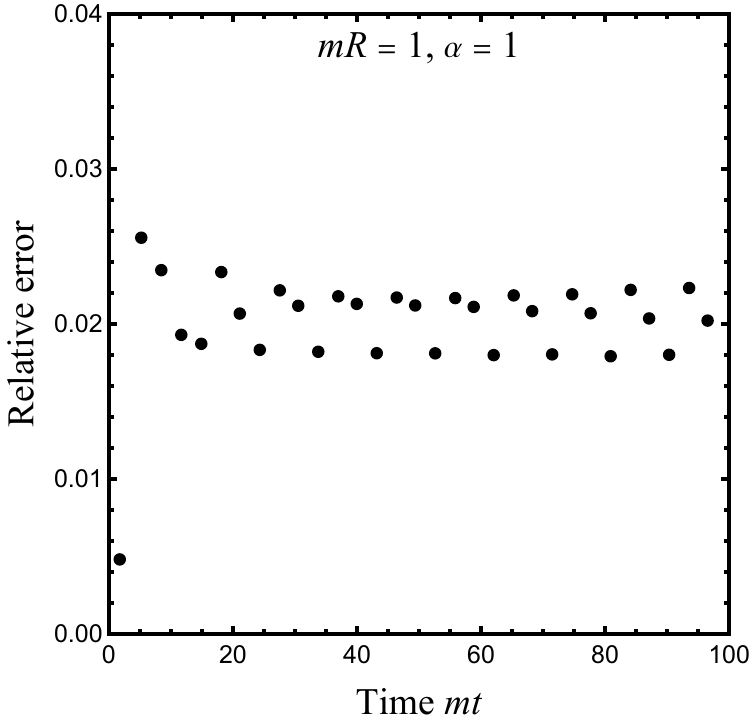}
    \caption{\small {\bf Left:} Comparison between numerical and semi-analytic solutions, for a source mass of size and density $mR = \alpha = 1$. The colored curves represent the numerical solutions while the black points represent the semi-analytical approach. Comparisons are made for $\varphi(r, t)$ at $m r = 0, 1,~\mathrm{and}~5$. Note that we have added a constant to the $m r = 1,~\mathrm{and}~5$ cases in order to show everything more clear in the plot. This means that for $mr=1$ we are representing $\varphi+0.15$ and for $mr=5$ we are showing $\varphi+0.30$. {\bf Right:} Also shown is the error between the two approaches, for the specific case of $m r = 1$.  We see that the two approaches agree within $\approx 3\%$.}
    \label{fig:numeric-semianalytic-comparison}
\end{figure}

Our main results are plotted in Figs.~\ref{fig:amplitude-comparison} and \ref{fig:numeric-semianalytic-comparison} for the range of parameters that are accessible to numerical simulations, $0.1 < mR < 10$ and $0.1 < \alpha \lesssim 3$.  In each plot we show the amplitude of the oscillatory solutions, as explained in the previous paragraph.  This is repeated for multiple radii around each source. Agreement between the semi-analytical (points) and numerical (lines) results is excellent.
Importantly, we can see that the analytical estimate for the long-time behavior, Eq.~\eqref{sacada}, is an excellent match for field solutions outside the source.  Deep inside a large source the match is relatively poor, but in all cases the power law is roughly $t^{-1/2}$.

\subsection{Time taken for field to begin power law decay}

So far we have seen that for an appearing source, the fields tends to the stationary solution on a time-scale $(mt)^{-1/2}$. With this in mind, we might wish to measure how long it takes the field to approach the power law decay towards the stationary solution.  We can see from Fig.~\ref{fig:amplitude-comparison} that, especially for $mr = 10$, there are some initial oscillations that eventually settle into the $t^{-1/2}$ power law decay.  How long does that initial process take?  In the interest of answering this question we have performed simulations with additional values of $\alpha$.

\begin{figure}
    \centering

    \begin{tikzpicture}
\node at (0, 4.75*2){
\includegraphics[width=0.32\linewidth]{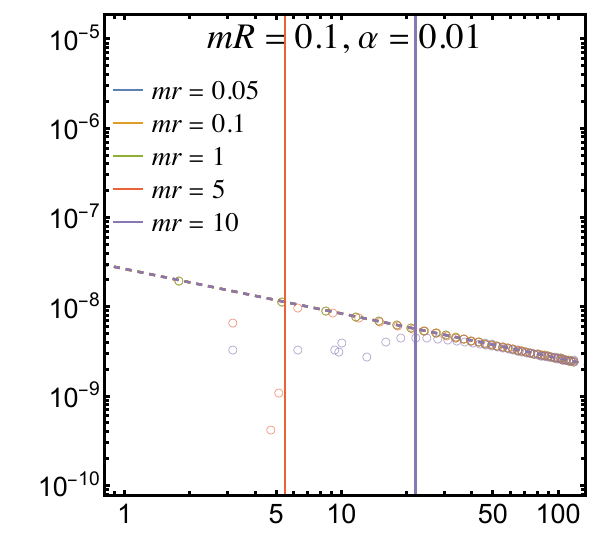}
\includegraphics[width=0.32\linewidth]{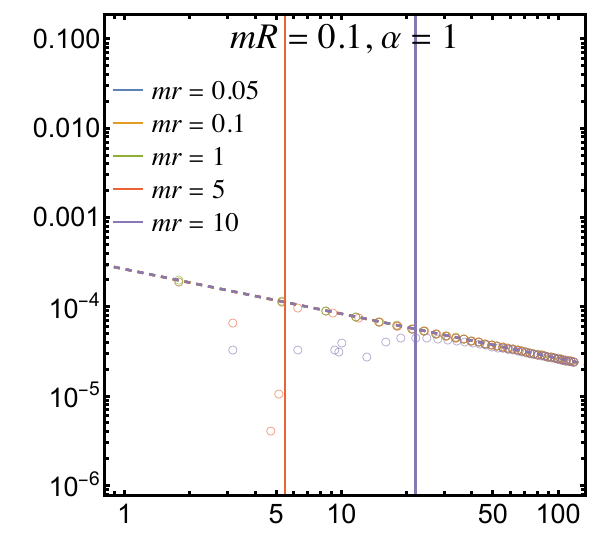}
\includegraphics[width=0.32\linewidth]{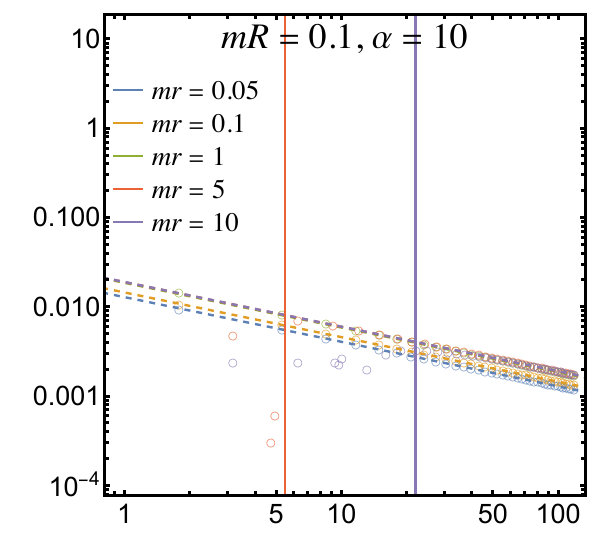}};

\node at (0, 4.75){
\includegraphics[width=0.32\linewidth]{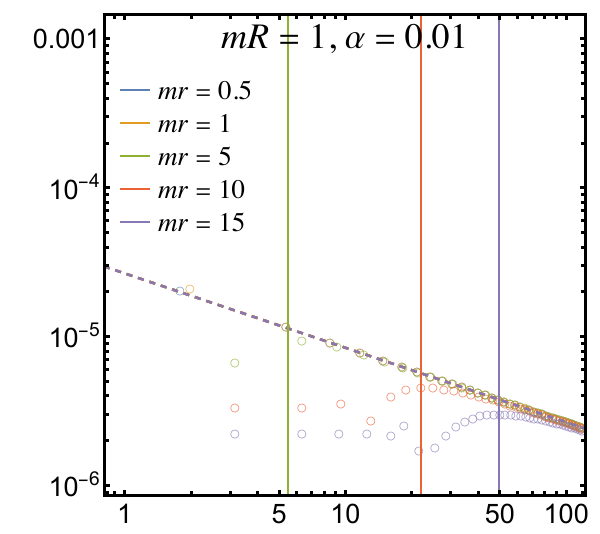}
\includegraphics[width=0.32\linewidth]{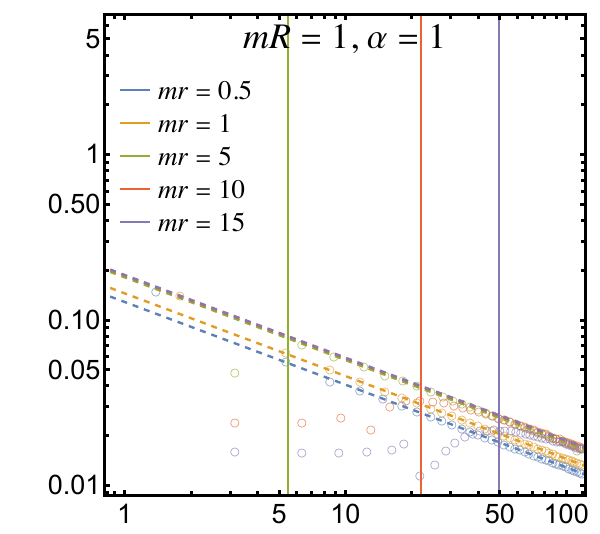}
\includegraphics[width=0.32\linewidth]{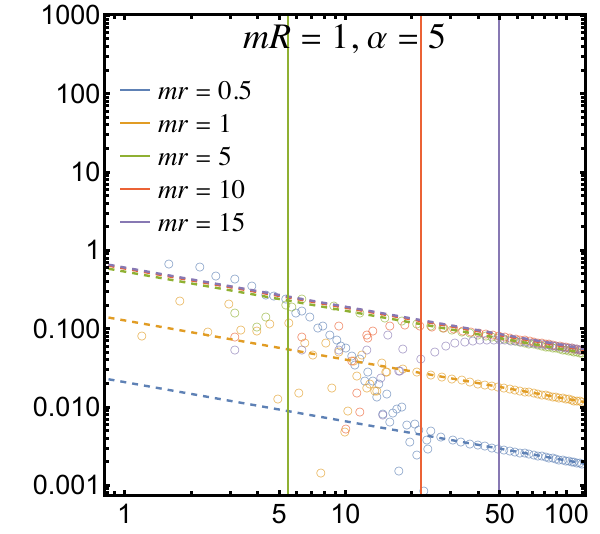}};

\node at (0,0){
\includegraphics[width=0.32\linewidth]{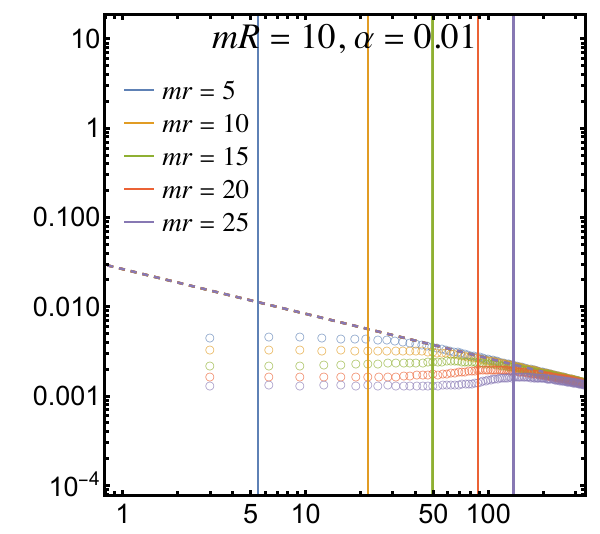}
\includegraphics[width=0.32\linewidth]{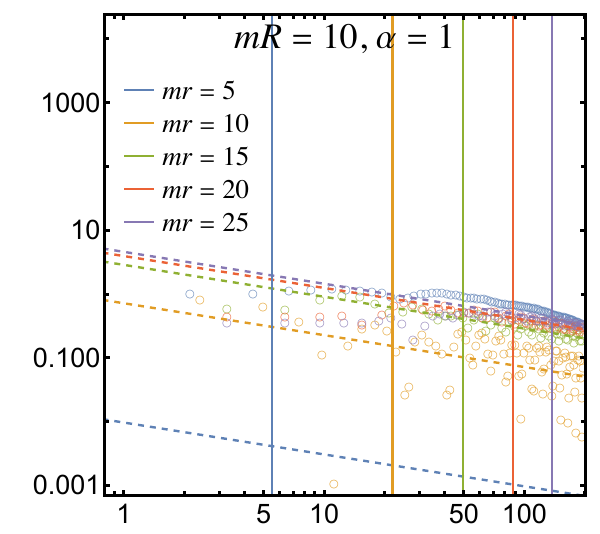}
\includegraphics[width=0.32\linewidth]{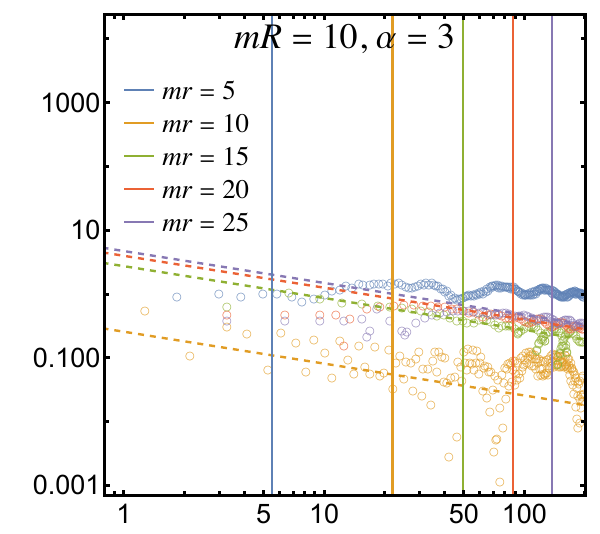}};

\node at (-4.75,-2.5){Time $m t$};
\node at (0.5,-2.5){Time $m t$};
\node at (5.75,-2.5){Time $m t$};

\node[rotate=90] at (-7.75, 0){Osc. amplitude of $\varphi$};
\node[rotate=90] at (-7.9, 4.75){Osc. amplitude of $\varphi$};
\node[rotate=90] at (-7.9, 4.75*2){Osc. amplitude of $\varphi$};
    
\end{tikzpicture}
    
   \caption{\small Results in the semi-analytical approach justifying that the approach to the long-time behavior, Eq.~\eqref{sacada}, occurs at the time scale given by Eq.~\eqref{eq:fitformula}. In each plot the solid, sloped lines show the behavior given in Eq.~\eqref{sacada}. The circles show the semi-analytical solutions. They approach the solid lines i.e. the late-time behavior, on a time scale given by Eq.~\eqref{eq:fitformula} and indicated by the vertical lines. We have omitted the values given by the fitting formula for $mt \lesssim 1 $ since the field has not yet made a whole oscillation and in the cases of interest, the field behavior is already well approximated by Eq.~\eqref{sacada}.}
    \label{fig:fittingformula}
\end{figure}

The solutions are shown in Fig.~\ref{fig:fittingformula}. Based on these simulations we have deduced a fitting formula that expresses the time that takes for the field to begin uniform power law decay, which is given by
\begin{equation}
    t\approx 0.22 mr^2~.
    \label{eq:fitformula}
\end{equation}
This expression is represented in Fig.~\ref{fig:fittingformula} as vertical lines. It can be seen that the fitting formula works quite well in almost all cases, accurately describing the times where the field behavior is driven by the power law decay. Note however, that this expression fails for $mr<mR$. 

\subsection{Summary of results}
\label{sec:timedependece-main}
Our results may be summarised in the following way.  For a given radius $r$ from the source mass, the evolution continues in lockstep with the homogeneous background oscillation until time $t = r - R$ as dictated by causality. 
From there, on a time scale of $\sim 0.22 mr^2$ the field approaches a long-time behavior that is well-described by Eq.~\eqref{sacada}. Notably during this late phase 
the field adjusts towards the stationary solution with an approach $\sim(m t)^{-1/2}$.   This behavior is observed across a wide range of source radii $0.1 < m R < 10$ and source densities $0.1 < \alpha < 3$.  In all cases excellent agreement is found between the numerical and semi-analytic methods.  The field's behavior is well approximated by Eq.~\eqref{sacada}, except deep inside large and dense sources. 
That said, our main interest is this work is the phenomenology of the field at or above the surface of the Earth, where Eq.~\eqref{sacada} performs well.

\section{ Energy and Momentum Flow}\label{sec:enermom}

Having shown that the numerical and semi-analytic approaches yield equivalent field solutions, we now turn to an additional consistency check that also serves to understand the questions raised in Sec.~\ref{sec:model}: we would expect that our solutions conserve energy and that the infinite energy differences between stationary solutions, observed in Sec.~\ref{sec:model} are resolved by a gradual energy outflow over time.  

\begin{figure}
    \centering
    \includegraphics[width=0.48\textwidth]{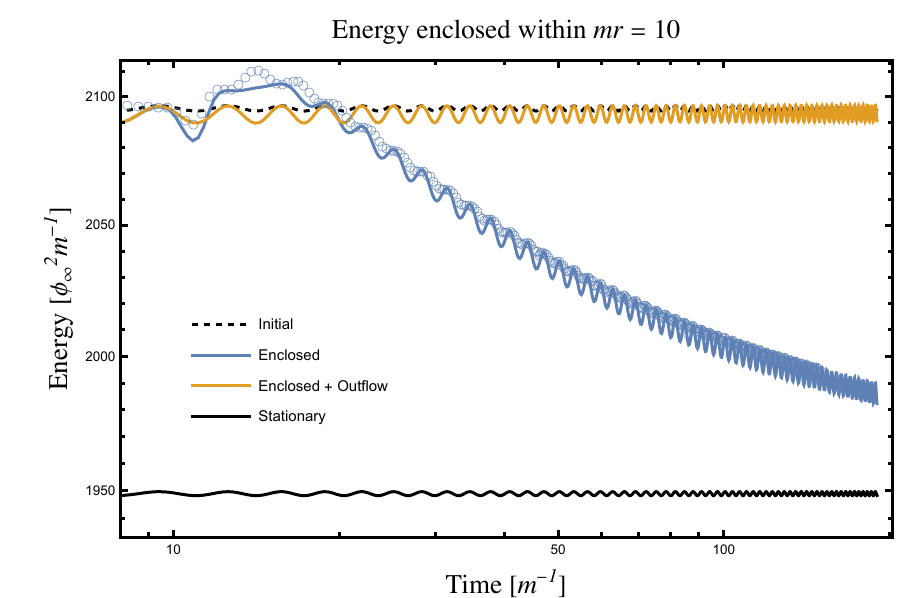}
    \hspace*{0.4cm}\includegraphics[width=0.48\textwidth]{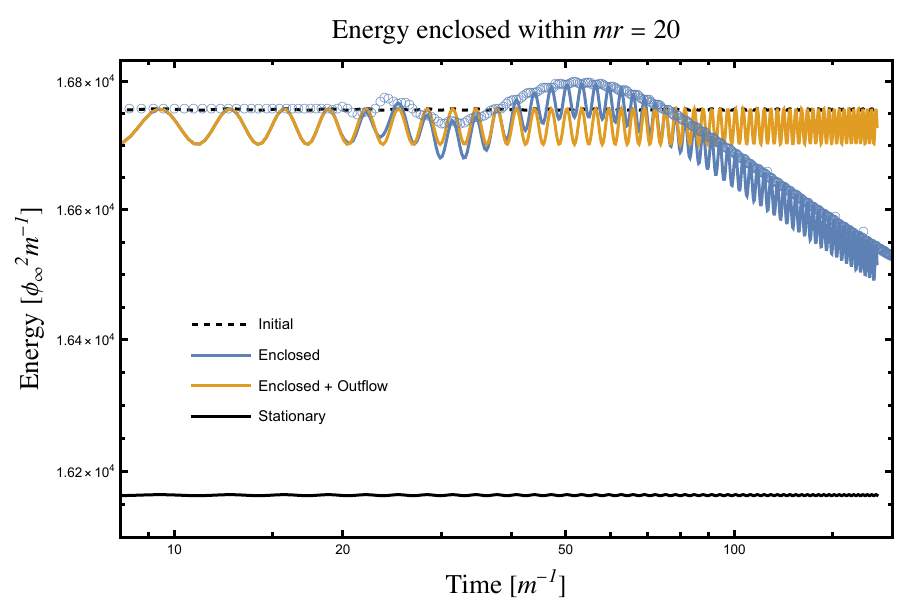}
    \caption{\small Measured energy outflow through a radius $m r = 10$ (left) and $20$ (right).  A source of $m R = \alpha = 1$ appears at $t = 0$, causing a subsequent expulsion of energy. {The energy enclosed is represented as a solid blue line for the numerical approach and as blue circles for the semi-analytic one.}  It is seen that the energy within this radius converges towards that of the stationary field configuration Eq.~\eqref{phi-solution-quasistatic}.  It is also seen that the energy contained within the radius, plus the integrated energy outflow (defined as the time integral of Eq.~\eqref{energy-outflow} at $mr = 10$ or $20$) remains constant, confirming that the energy flows outwards and serving as a useful check of the accuracy of the numerical procedure.}
    \label{fig:energy-outflow}
\end{figure}

The results for an appearing source with $m R = \alpha = 1$ are depicted in Fig.~\ref{fig:energy-outflow}. It can be seen that the effect of the appearing (repulsive) source is to expel  energy from its vicinity,  where the energy density is defined as the $T_{00}$ component of the field's energy-momentum tensor:
\begin{equation}
    T_{\mu \nu} = \eta_{\mu \nu} {\cal L} + \partial_\mu \phi \partial_\nu \phi~.
    \label{energymomentumtensor}
\end{equation}
The Lagrangian here is the integrand of Eq.~\eqref{action}.  The enclosed energy within a radius $R_{\rm{sphere}}$ is then given by,
\begin{equation}
    E(t) = \int_0^{R_{\rm{sphere}}} 4 \pi r^2 T_{00}(t, r) dr~.
\end{equation}
We see that the energy in the vicinity gradually converges to that of the stationary solution Eq.~\eqref{phi-solution-quasistatic}.
The instantaneous energy outflow through the boundary at $R_{\rm{sphere}}$ is defined as
\begin{equation}
    \dot E(t) = 4 \pi (R_{\rm{sphere}})^2 T_{0 i}(t, r = R_{\rm{sphere}})~.
    \label{energy-outflow}
\end{equation}
We also see that the energy contained within $R_{\rm{sphere}}$, plus the energy that flows outwards (computed as the time integral of Eq.~\eqref{energy-outflow}) remains constant, as one would expect.  Seeing that the numerical solutions conserve energy is a useful check that the numerical method is well-behaved.

Comparing the curves in Fig.~\ref{fig:energy-outflow} for the two different values for $R_{\rm{sphere}}$ we can clearly see that the required time to converge towards the energy of the stationary solution increases with increasing radius. In this sense we can understand that the stationary solution is approached over time as energy is expelled from larger and larger radii.

\section{Discussion of experimental time-scales}\label{sec:timeresults}

As already mentioned, the time scale for the approach to the stationary solution will be particularly relevant for accelerated motion. 
To get an impression of its importance let us consider the time-scales on which acceleration matters for the case of Earth orbiting the Sun.

For experiments taking place in the vicinity of Earth's surface or in a low orbit we can consider the time it takes for the Earth's orbit to deviate from linear motion by an amount equal to the size of the Earth itself. 
Assuming a circular orbit taking time $T_{\rm orbit}$ to complete one orbit of radius $R_{\rm orbit}$ the distance from the straight path is given for small angles $\gamma$ by,
\begin{equation}
    d=R_{\rm orbit}\frac{\gamma^2}{2}~.
\end{equation}
This distance is reached after a time,
\begin{equation}
\label{eq:earthmove}
    t=\frac{T_{\rm orbit}}{2\pi}\gamma=\sqrt{\frac{2d}{R_{{\rm orbit}}}}\frac{T_{\rm orbit}}{2\pi}\sim 0.8\,{\rm days}\left(\frac{d}{2\,R_{\rm Earth}}\right)^{1/2}.
\end{equation}

\bigskip

In the following we will base our analysis on the time scales obtained in Sec.~\ref{sec:timedependence}, in particular Eqs.~\eqref{sacada}and \eqref{eq:fitformula}. Due to limited computational power they have been obtained in a regime of moderate values for $\alpha$ and $mR$. While we observe a reasonably consistent behavior over this range we should nevertheless keep this caveat in mind when we extrapolate these results to values much different from 1, as we will do below.
\bigskip

\begin{figure}
\centering
\includegraphics[width=0.5\textwidth]{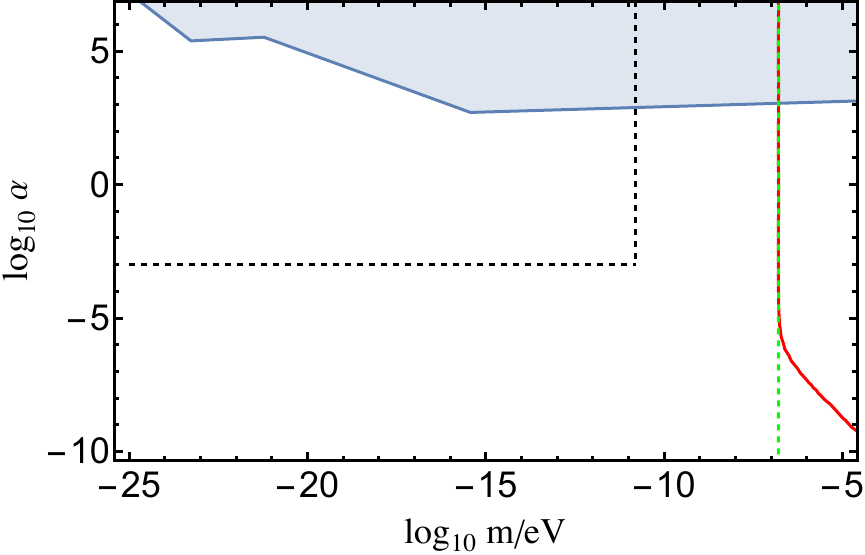}
      \caption{\small Region where the relaxation to the stationary solution is faster than the time scale for Earth to deviate by one Earth radius from a straight path (left of red line). In the region to the left of the green dashed line, the time evolution is also fast for the spatially dependent part of the field on the surface of the Earth. {For a rough comparison the blue region shows the limits obtained in Ref.~\cite{Hees:2018fpg} for their combination $d^{(2)}_{\hat{m}}-d^{(2)}_{g}$, of average light quark mass and gluon coupling, respectively, but assuming that only the gluon coupling is non-zero. We have taken this to be the sole non-vanishing coupling. Moreover, we only considered the leading non-composition dependent part and used an average density for Earth of $5.51{\rm g}/{\rm{cm}}^3$~\cite{wikiearth}. }
      {In the region to the upper left of the black dashed line effects of a non-vanishing velocity of the dark matter background can be neglected. The horizontal line compares the non-relativistic potential to the kinetic energy of the dark matter particles and the vertical one compares the deBroglie wavelength to the diameter of Earth. For both we have assumed $v_{\rm DM}\sim 10^{-3}$.} }
      \label{fig:compare}
\end{figure}

We are mostly interested in the regime of very small masses that is most tightly constrained by considerations of long-range forces.
Moreover, for larger dark matter particle masses the non-vanishing velocity $v_{\rm DM}$ of dark matter needs to be taken into account, which is beyond the scope of the present paper. There are two effects related to this. First, when the de Broglie wave-length of the dark matter particles $\sim 1/(mv_{\rm DM})$ becomes of the order of the size of Earth or smaller the spherically symmetric treatment may be questionable. This typically occurs when $m\ll 10^{-10}\,{\rm eV}$. Second, the kinetic energy $\sim mv^{2}_{\rm DM}$ of the dark matter particles may allow them to penetrate the barrier provided by the repulsive potential caused by the quadratic interaction with matter, changing the solutions.   This is relevant when $V_{\rm barrier}\sim \alpha^2 m\lesssim mv^{2}_{\rm DM}$, i.e. $\alpha\lesssim 10^{-3}$ (see also Ref.~\cite{yeray}). As we will see below and in Fig.~\ref{fig:compare} (where these conditions are indicated by the dashed black lines) we find the time evolution to be fast in the region where we can neglect the velocity $v_{\rm{DM}}$.

\bigskip

According to the fit formula Eq.~\eqref{eq:fitformula} the approach to the late-time solution Eq.~\eqref{sacada} occurs on the timescale
\begin{equation}
    t\sim 0.22 mr^2\lesssim 0.8\,{\rm days}\left(\frac{m}{10^{-7}\,{\rm eV}}\right)\left(\frac{r}{R_{Earth}}\right)^2.
\end{equation}
In the low-mass regime we can therefore safely assume that the approach to Eq.~\eqref{sacada} is very fast compared to the timescale of the Earth's motion.
The (relative) deviation from the stationary solution is therefore controlled by the size of the dimensionless quantity
\begin{equation}
\label{eq:scaleevolve}
\left(\frac{\alpha m R-\tanh(\alpha m R)}{\alpha} \right)\sqrt{\frac{2}{\pi m t}}~,
\end{equation}
relative to 1.
Depending on the coupling strength and the size of the source object we have two regimes depending on the size of $\alpha mR$. In particular the late-time solution Eq.~\eqref{sacada} is quite close to the stationary solution when
\begin{eqnarray}
\label{eq:small}
t&\sim&\frac{2}{9\pi m}\frac{(\alpha mR)^6}{\alpha^2}\sim 0.06\,{\rm days}\left(\frac{10^{-20}{\rm eV}}{m}\right) \frac{(\alpha mR)^6}{\alpha^2}\qquad\qquad\qquad~\alpha mR\ll 1~,\\
\label{eq:large}
t&\sim& \frac{2}{\pi}mR^2\sim 0.6 m R^2\sim 2\,{\rm days}\left(\frac{m}{10^{-7}\,{\rm eV}}\right)\left(\frac{R}{R_{\rm Earth}}\right)^2\qquad\qquad\quad\, \alpha mR\gg 1~.
\end{eqnarray}

The {region} where these timescales are comparable to those of Earth's motion are shown in Fig.~\ref{fig:compare}. The filled region shows the limits obtained in~\cite{Hees:2018fpg} for a specific combination of a quadratic coupling to electrons and gluons (see~\cite{Hees:2018fpg} for details).  The region left of the red line features a fast time evolution according to Eq.~\eqref{eq:scaleevolve}.

\bigskip

Comparing to the full field values of the stationary solution is most relevant when considering experiments that are directly sensitive to the amplitude/density of the field.
However, some experiments, notably the ones discussed in~\cite{Hees:2018fpg} but also, e.g.~CASPER-wind~\cite{JacksonKimball:2017elr} and experiments using accelerometers~\cite{Graham:2015ifn}, are mainly sensitive to gradients in the field.\footnote{We note that the two latter experiments are based on a linear coupling to matter. But in presence of a quadratic coupling that changes the field profiles their results need to be reinterpreted, cf.~\cite{Banerjee:2022sqg}} In this case we should consider only the space-dependent part of the stationary solution as a reference.

Let {us therefore} consider Eq.~\eqref{phi-solution-quasistatic} outside the source object, i.e. at $r>R$ and focus on the non-homogeneous part, 
\begin{equation}
    \phi_\mathrm{st}(r, t)-\phi_{\infty} = - \phi_{\infty}\cos(m t + \delta )            \left(\frac{\alpha m R - \tanh \alpha m R}{\alpha m R}\right)\frac{R}{r} \quad{\rm for}\quad r \geq R~.
\end{equation}
Let us now consider the relative deviation in this space-dependent part,
\begin{equation}
    \frac{\phi_{0}(r,t)-\phi_{\infty}}{\phi_\mathrm{st}(r, t)-\phi_{\infty}}\approx (mR)\sqrt{\frac{2}{\pi mt}}\frac{r}{R}~.
\end{equation}
As it turns out the timescale is again independent of $\alpha$,
\begin{equation}
\label{eq:timeneeded}
    t\sim \frac{2}{\pi}mr^2\sim 0.6 m R^2\sim 2\,{\rm days}\left(\frac{m}{10^{-7}\,{\rm eV}}\right)\left(\frac{r}{R_{\rm Earth}}\right)^2\,.
\end{equation}
This area is indicated by the dashed green line in Fig.~\ref{fig:compare}.
The red and the green dashed line agree in their vertical part because we have chosen the Earth radius for the location of the experiment.

\bigskip

Test particles are also not always stationary either in the reference frame of the background dark matter, or in the frame  of the Earth.  We take as an example the MICROSCOPE satellite which obtained stringent constraints on violations of the weak equivalence principle \cite{MICROSCOPE:2022doy}, which has also been used to constrain quadratically coupled dark matter \cite{Hees:2018fpg}. The satellite completed 1960 orbits of the Earth in 135 days, and housed test masses of size $\sim 5 \mbox{ cm}$ within a satellite of size 
$\sim 2 \mbox{ m}$.  As the satellite moves at a speed of $\sim 8 \mbox{ cm/s}$ the satellite moves a distance of order its own size in a time of $\sim 3 \times 10^{-3}\mbox{ s}$.  The longest response times of the field are in the $\alpha mR \gg 1$ regime (which is valid for the satellite across the majority of our parameter space of interest), and this response time is 
\begin{equation}
    t \lesssim 4 \times 10^{-2} \left( \frac{m}{\mbox{eV}}\right) \rm{s}~.
\end{equation}
We conclude that despite the rapid motion of the satellite, its relatively small size leads to rapid response times in the parameter space of interest, and thus also in this case the stationary approximation holds.

\section{Conclusions}\label{sec:conclusions}
Very light bosonic dark matter that couples quadratically to matter (e.g. nucleons) exhibits an interesting ``environmental'' dependence in the vicinity of matter~\cite{Hees:2018fpg,Banerjee:2022sqg}. In this model the density of dark matter is modified and exhibits a non-trivial profile around compact sources such as the Earth. This directly affects attempts to directly detect such forms of dark matter~\cite{Banerjee:2022sqg} but can also lead to new tests, e.g. via oscillating forces~\cite{Hees:2018fpg}.

Comparing stationary solutions in the presence and absence of matter, however, some peculiarities arise. Notably, the energy difference between e.g. a solution in presence of a spherical object like Earth and in its absence, feature an infinite difference in the total energy, arising at large radii/volumes.  Even worse, such infinities are present even when just changing the size of the object in question. As discussed in this note, these infinities 
hint at the relevance of the time dependence of the problem. As we go further and further away from the source it takes a longer and longer time to approach the stationary solution, as more energy needs to be expelled from (or attracted into) the relevant volume (cf. Eq.~\eqref{eq:timeneeded}).

This also raises the question: under which circumstances are the stationary solutions applicable? Earth moves on an accelerated trajectory, so the field configuration that it sources should differ from the stationary solution.
We are therefore interested in how quickly such solutions are approached.
To this end we have studied and quantified the required times by both semi-analytical as well as fully numerical calculations in a simplistic situation where the relevant object simply ``appears''. 
Outside the source object the stationary solutions are approached with a behavior that can be well quantified by the approximate late-time expression Eq.~\eqref{sacada}.
We have argued that for most of the mass range of interest, the approach to the stationary solutions is quite fast compared to the relevant timescales and for the Earth we can safely use the stationary solutions.  However, we emphasize that in this work we have relied on a simplified scenario in which the time dependence of the source is modelled by a sudden appearance at time $t = 0$.  It would still be of interest to obtain solutions with a setup that more accurately models the dynamics of the Solar System, particularly the motion of the Earth around the Sun.  Likewise, our numerical approach is limited to particular mass and coupling scales, so it is difficult to make completely generic statements about these effects.  Nevertheless, we believe that our setup has captured at least the leading-order effects of {the} time dependence in these systems.

\subsection*{Acknowledgements}
BE is supported in part by the STFC Consolidated Grants ST/T000791/1 and ST/X000575/1, as well as Simons Investigator award 690508.
JJ gratefully acknowledges support by an IPPP DIVA fellowship and a European Union’s Horizon 2020 Marie Sklodowska-Curie grant (No 860881-HIDDeN). CB is supported by  STFC Consolidated Grant [Grant No. ST/T000732/1]. Part of this work was done in the context of YGCs master thesis and is based on it. For the purpose of open access, the
authors have applied a CC BY public copyright licence to any Author Accepted Manuscript
version arising.

\subsection*{Data Availability Statement} This work is entirely theoretical and has no associated
data.

\appendix

\section{Contracting and expanding source}
\label{app:contracting}
{As mentioned in the main text, a still simple but perhaps somewhat more physical situation (compared to the appearing source)} is a source that suddenly contracts or expands while conserving its mass. 

\begin{figure}[t]
   
        \centering
    \includegraphics[width=0.45\textwidth]{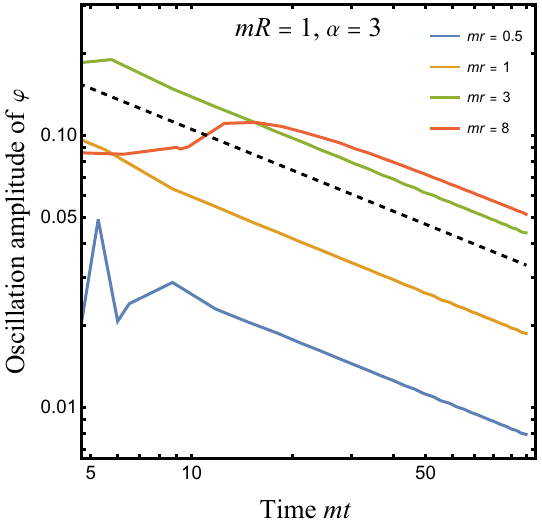}
    \hspace{1cm}
    \includegraphics[width=0.45\textwidth]{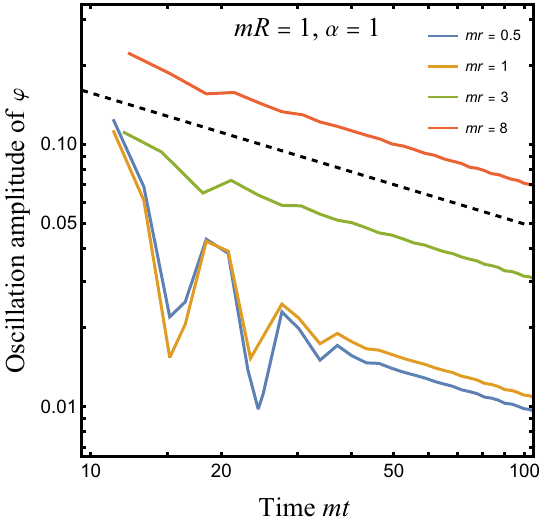}
    \caption{\small {\bf Left:} Time scaling for the contracting source for $R/R_0=1/3$, $\alpha=3$ and $mR=1$. {\bf Right:} Time scaling for the expanding source for $R/R_0=3$, $\alpha=1$ and $mR=1$.  The dashed line represents a decaying behavior $\sim \frac{1}{\sqrt{t}}$.}
  \label{semicon}
\end{figure}

\bigskip
In order to investigate the situation of a contracting or expanding source, we can again use the semi-analytical approach. The initial field distribution is given by the stationary solution for the initial radius of the source $R_0$. We can project this initial distribution on the complete set of eigenfunctions of the new-sized source with radius $R$, cf. Appendix~\ref{eigenfunctions}. Since the time evolution for the eigenfunctions is known, the time evolution for the initial configuration can be easily computed. As in the appearing source case, we expect that the solution will tend towards the stationary solution $\phi_\mathrm{st}$ of the re-sized source as $t\to \infty$. As can be seen from Fig.~\ref{semicon}, the time scaling behavior is the same one we found for the disappearing and appearing source, $\sim \frac{1}{\sqrt{t}}$.

\begin{figure}[t]
   
        \centering
         \centering
    \includegraphics[width=0.48\textwidth]{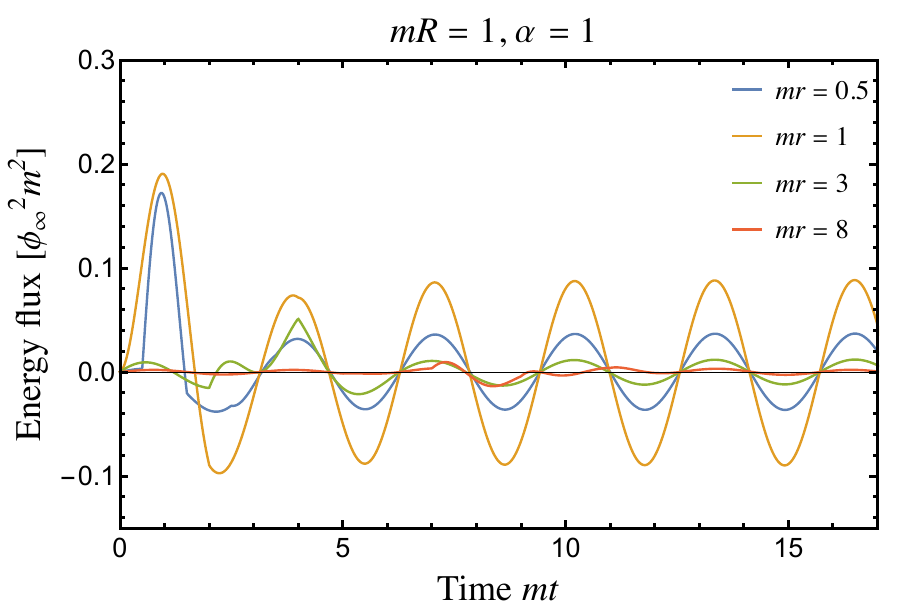}
    \hspace*{0.4cm}\includegraphics[width=0.48\textwidth]{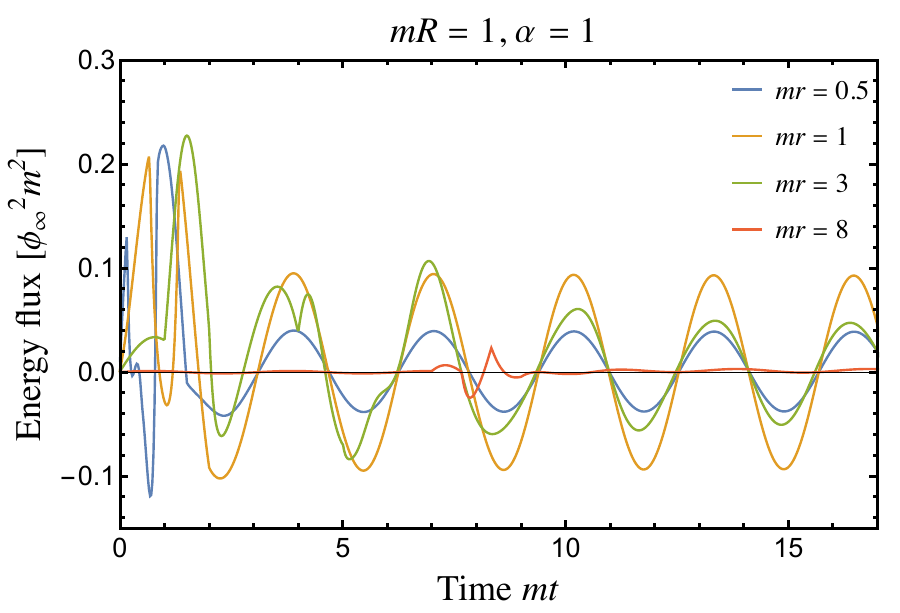}    
    \caption{\small {\bf Left:} Energy flow for the contracting source for $R/R_0=1/3$, $\alpha=1$ and $mR=1$. {\bf Right:} Energy flow for the expanding source for $R/R_0=3$, $\alpha=1$ and $mR=1$.}
  \label{enflow}
\end{figure}

We can also compute the energy flow through an infinitesimal spherical surface using the $0r$ component of the energy-momentum tensor given by Eq.~\eqref{energymomentumtensor}. As we can see in Fig.~\ref{enflow},  although in an infinite volume there is an infinite energy difference between the stationary configurations before and after the change of size, for a finite surface/region the energy flow remains finite at all times. Note that the field relaxes to an oscillatory stationary behavior. This was expected, since the field tends to the new stationary solution of the re-sized source.

\section{Computation of the eigenfunctions of the differential equation}
\label{eigenfunctions}

In this appendix we compute the  set of eigenfunctions for the problem of a spherical source in an oscillating background. We also provide the appropriate normalizations.

In order to find the eigenfunctions of the differential equation 
\begin{equation}
(\partial_{t}^{2}-\nabla^{2} +m^{2}+ \frac{\rho}{M^2}) \phi=0 ~,
\label{eqdif}
\end{equation}
with
\begin{equation}
    \rho(\vec{x})=\rho\Theta(R-|\vec{x}|)~,
\end{equation}
we make the ansatz $\phi(\vec{r},t)=\Tilde{\phi}(\vec{r})e^{-i \omega t}$. The solutions for $\Tilde{\phi}(\vec{r})$ will be the eigenfunctions associated with the different frequencies. Plugging the ansatz into the differential equation, the Helmholtz or the modified Helmholtz equation is obtained depending on the parameters.  
\paragraph{}
For $\omega^2 -m^2 -\frac{\rho}{M^2} > 0$, the eigenfunctions read
       \begin{equation}
    \Tilde{\phi}_{k,l,m,\pm }(\vec{r})= 
     \begin{cases}
  C_{k,l,m,\pm}  j_l(k' r)Y_{l,m}(\theta,\varphi)& \text{if } r<R~,\\
      (A_{k,l,m,\pm} h_l^1(kr)+B_{k,l,m,\pm} h_l^2(kr))Y_{l,m}(\theta,\varphi)& \text{if }  r>R~,\\
       \end{cases}
   \end{equation}
where $\pm$ indicates the sign of $\omega$, $k'=\sqrt{\omega^2-m^2-\frac{\rho}{M^2}}$ and $\omega= \pm \sqrt{m^2+k^2}$. The constants have to be determined with the continuity of the field and its derivative at the radius $r = R$, as well as the normalization of the eigenfunctions.

        For $\omega^2 -m^2 - \frac{\rho}{M^2}< 0$, the eigenfunctions read
      \begin{equation}
    \Tilde{\phi}_{k,l,m,\pm }(\vec{r})= 
     \begin{cases}
 C_{k,l,m,\pm} i_l(k' r)Y_{l,m}(\theta,\varphi)& \text{if } r<R~,\\
    (A_{k,l,m,\pm} h_l^1(kr)+B_{k,l,m,\pm} h_l^2(kr))Y_{l,m}(\theta,\varphi)& \text{if }  r>R~,\\
       \end{cases}
        \end{equation} 
 where in this case $k'=\sqrt{-\omega^2+m^2+\frac{\rho}{M^2}}$ and again the constants have to be found. 
 \paragraph{}
 Since the differential equation is quadratic in time, two conditions have to be imposed on the problem. We require the field to be real and $\frac{\partial \phi(\vec{r},0)}{\partial t}=0$, in this way, the eigenfunctions will fulfill $\phi(\vec{r},t)=\Tilde{\phi}(\vec{r})\cos( \omega t)$. The situations that will be treated in this work have spherical symmetry, so basically the only relevant eigenfunctions will be the ones with $l=m=0$, given by 
   \begin{equation}
   \Tilde{\phi}_{k}(\vec{r})= 
     \begin{cases}
  C  j_0(k' r)& \text{if } r<R~,\\
    A h_0^1(kr)+B h_0^2(kr)& \text{if }  r>R~,\\
       \end{cases}
       \label{eigen1}
         \end{equation}
for $\omega^2 -m^2 -\frac{\rho}{M^2} > 0$ and 
   \begin{equation}
  \Tilde{ \phi}_{k }(\vec{r})= 
     \begin{cases}
  C i_0(k' r)& \text{if } r<R~,\\
     A h_0^1(kr)+B h_0^2(kr)& \text{if }  r>R~,\\
       \end{cases}
       \label{eigen2}
        \end{equation}
       for $\omega^2 -m^2 -\frac{\rho}{M^2} < 0$.
       \paragraph{}
Now, three conditions are required to fix the remaining constants. Two of them come from the continuity of the field and its derivative at the radius. 

\subsubsection*{Case $\mathbf{\omega^2 -m^2 -\frac{\rho}{M^2} > 0}$}
For $\omega^2 -m^2 -\frac{\rho}{M^2} > 0$ the continuity conditions read,
\begin{equation}
     \begin{cases}
C\frac{\sin(k'R)}{k'R} =-\frac{iAe^{ikR}}{kR}+\frac{iBe^{-ikR}}{kR}~,\\
 C(\frac{\cos(k'R)}{R}-\frac{\sin(k'R)}{k'R^2})=Ae^{ikR}(\frac{1}{R}+\frac{i}{kR^2})+Be^{-ikR}(\frac{1}{R}-\frac{i}{kR^2})~,\\
       \end{cases}
        \end{equation}
       which can be solved for $A$ and $B$
 \begin{equation}
       A=\frac{Ce^{-ikR}}{2}\left(\cos(k'R)+\frac{ik\sin(k'R)}{k'}\right)~,
        \end{equation}
       \begin{equation}  
   B=\frac{Ce^{ikR}}{2}\left(\cos(k'R)-\frac{ik\sin(k'R)}{k'}\right)~.
   \end{equation}
Now, in order to fix the remaining constant $C$, a third condition is required. In principle, it is preferred to have an orthonormal basis since it is easier to project on it. Since the basis we are considering is continuous, we require that the basis is orthonormal in a Dirac delta sense

\begin{equation}\int \overline {\Tilde{\phi}_{k_1 }(\vec{r})} \Tilde{\phi}_{k_2}(\vec{r}) \,dV=\delta^3(\vec{k_1}-\vec{k_2})~,
\end{equation}
in order to fix the remaining constant. The scalar product of two eigenfunctions reads

\begin{eqnarray}
\int \overline {\Tilde{\phi}_{k _1}(\vec{r})} \Tilde{ \phi}_{k_2 }(\vec{r}) \,dV&&
\\\nonumber
&&\!\!\!\!\!\!\!\!\!\!\!\!\!\!\!\!\!\!\!\!\!\!\!\!\!\!\!\!\!\!\!\!\!\!\!\!\!\!\!\!\!\!\!\!\!\!\!=
4\pi \overline {C_1}C_2\int_0^R r^2 j_0(k_1'r) j_0(k_2'r) \,dr
\\\nonumber
&&\!\!\!\!\!\!\!\!\!\!\!\!\!\!\!\!\!\!\!\!\!\!\!\!\!\!\!\!\!\!+4\pi \int_R^\infty r^2 \overline{(A_{1} h_0^1(k_1r)+B_{1} h_0^2(k_1r))} (A_{2} h_0^1(k_2r)+B_{2} h_0^2(k_2r)) \,dr~,
\\\nonumber
&&\!\!\!\!\!\!\!\!\!\!\!\!\!\!\!\!\!\!\!\!\!\!\!\!\!\!\!\!\!\!\!\!\!\!\!\!\!\!\!\!\!\!\!\!\!\!\!=-4\pi\overline {C_1}C_2\frac{(k_2'-k_1')\sin((k_1'+k_2')R)-(k_2'+k_1')\sin((k_1'-k_2')R)}{2k_1'k_2'(k_2'^2-k_1'^2)}
\\\nonumber
&&\!\!\!\!\!\!\!\!\!\!\!\!\!\!\!\!\!\!\!\!\!\!\!\!\!\!\!\!\!\!+\frac{4\pi}{k_1k_2}\int_R^\infty \overline{A_1} A_2 e^{i(k_2-k_1)r}+\overline{B_1} B_2 e^{i(k_1-k_2)r} - \overline{A_1} B_2 e^{-i(k_2+k_1)r}-\overline{B_1} A_2 e^{i(k_1+k_2)r}  \,dr~.
\end{eqnarray}

At this point, we can use that $\int_0^\infty e^{ikx}\,dx= \pi \delta(k)+\frac{i}{k}$, which implies that $\int_R^\infty e^{ikx}\,dx=e^{ikR}( \pi \delta(k)+\frac{i}{k})$ in the $r>R$ integral. This leads to
\begin{eqnarray}
\int \overline {\Tilde{\phi}_{k _1}(\vec{r})} \Tilde{ \phi}_{k_2 }(\vec{r}) \,dV
\!&=&\!-4\pi \overline {C_1}C_2 \frac{(k_2'-k_1')\sin((k_1'+k_2')R)-(k_2'+k_1')\sin((k_1'-k_2')R)}{2k_1'k_2'(k_2'^2-k_1'^2)}
\\\nonumber
&&\,\,\,\,+4\pi \overline {C_1}C_2 \frac{(k_2'-k_1')\sin((k_1'+k_2')R)-(k_2'+k_1')\sin((k_1'-k_2')R)}{2k_1'k_2'(k_2^2-k_1^2)}
\\\nonumber
&&\qquad+\frac{4\pi^2}{k_1^2}(|A_1|^2+|B_1|^2)\delta(k_1-k_2)~.
\end{eqnarray}
The first and second term cancel since $k_2'^2-k_1'^2=k_2^2-k_1^2$. With this we obtain,
\begin{equation}
C=(8 \pi^3 (\cos^2(k'R)+\frac{k^2}{k'^2}\sin^2(k'R))^{-\frac{1}{2}}~.
\end{equation}

\subsubsection*{Case $\mathbf{\omega^2 -m^2 -\frac{\rho}{M^2} < 0}$}
For $\omega^2 -m^2 + \frac{\rho_A}{M^2}< 0$, the computation is analogous and we find,
\begin{eqnarray}
 A&=&\frac{Ce^{-ikR}}{2}(\cosh(k'R)+\frac{ik\sinh(k'R)}{k'})~, 
 \\
 B&=&\frac{Ce^{ikR}}{2}(\cosh(k'R)-\frac{ik\sinh(k'R)}{k'})~, 
 \\
 C&=&(8 \pi^3 (\cosh^2(k'R)+\frac{k^2}{k'^2}\sinh^2(k'R))^{-\frac{1}{2}}~.
\end{eqnarray}

\section{Derivation of the asymptotic formula}
\label{app:asymptotic}
It is possible to derive an analytical expression for the asymptotic behavior of the field in the appearing source situation. In order to pursue this objective we will use the expression of the time evolution of the field given by Eq.~\eqref{418}.

We start by taking a look at the projection of the initial configuration. Using the expressions for the eigenfunctions from Eq.~\eqref{eigen2}, the projection reads
\begin{eqnarray} \nonumber
\braket{\Tilde{\phi}|\phi_0}\!\!\!\!\!&=&\!\!\!\!\!\int\overline {\Tilde{\phi}_k(\vec{r})}\phi_0 (\vec{r},0)\,dV~,
   \\ \nonumber
   \!\!\!\!\!&=&\!\!\!\!\!4\pi C(k)\int_0^R i_0(k'r) r^2 \,dr +\frac{4\pi \overline{A(k)} i}{k}\int_R^\infty e^{-ikr-\epsilon r} r \,dr-\frac{4\pi \overline{B(k)} i}{k}\int_R^\infty e^{ikr-\epsilon r} r \,dr~,
   \label{ar}
   \\
   \!\!\!\!\!&\approx&\!\!\!\!\!\frac{4\pi C(k) e^{-\epsilon R}}{(k^2+\epsilon^2)^2}\left(\!(k^2 \epsilon R \!+\!R\epsilon^3\!-\!k^2+\epsilon^2)\frac{\sinh(k'R)}{k'}\!+\!(k^2R\!+\!\epsilon^2 R\!+\!\epsilon)\cosh(k'R)\!\right)~.
   \label{puf}
\end{eqnarray}
Note that for long times, but still fulfilling, $\frac{1} {\epsilon}\gg t$, the eigenstates that will drive the behavior of the field are the ones close to $k=0$, since we know the field tends to the stationary configuration, which has vanishing momentum. Then the first integral in Eq.~\eqref{ar} in this regime can be ignored, since its contribution to the projection is much smaller compared to the second and third ones. Doing an expansion around $\epsilon=0$ in Eq.~\eqref{puf}, we can approximate the whole projection to

 \begin{equation}
\braket{\Tilde{\phi}_k|\phi_0}\approx \frac{4 \pi^2 C(k)}{k^2}\cosh(k'R)\delta_\epsilon(k)+ \frac{ 4 \pi C(k)}{k^2}\left(\frac{\cosh(k'R)k'R-\sinh(k'R)}{k'}\right)~,
\label{apr}
 \end{equation}
where $\delta_{\epsilon}(k)$ is a Dirac delta in the limit $\epsilon\rightarrow 0$. Strictly speaking additional delta-functions appear when doing the expansion of Eq.~\eqref{puf} but they are multiplied by higher powers of $k$ so they vanish upon evaluation of the integral. Note also that a factor $1/2$ arises in the first term of Eq.~(\ref{apr}), from the fact that the lower limit of the integral in spherical coordinates is at $k=0$. Plugging this result into the time evolution,   Eq.~(\ref{418}), the field reads
 \begin{multline}
 \label{eq:preapprox}
     \phi_0(\vec{r},t)\approx 4  \pi^2 \int \frac{ C(k)}{k^2} e^{-i\omega(k) t}\Tilde{\phi}_{k }(\vec{r})\cosh(k'R)\delta_{\epsilon}(k) \,dk^3
     \\
     +4 \pi\int \frac{C(k)}{k^2}e^{-i\omega(k) t}\Tilde{\phi}_{k }(\vec{r}) \left(\frac{\cosh(k'R)k'R-\sinh(k'R)}{k'}\right)\,dk^3~.
 \end{multline}
Note that, we have exchanged the $\cos(\omega(k))$ in Eq.~\eqref{418} for $e^{-i\omega(k) t}$. This simplifies the  computation and we just need to be careful of taking the real part of the resulting expression at the end. 

The first term in Eq.~\eqref{eq:preapprox} can be computed easily and the stationary solution arises. This is because for $k \rightarrow 0$ the eigenfunctions are given by
\begin{equation} 
\lim_{k\to 0} C(k) i_0(k'r)=\frac{1}{\sqrt{8 \pi^3}\cosh(\alpha mR)}\frac{\sinh(\alpha m r)}{\alpha m r}~,
\end{equation}
for $r<R$. For $r>R$ the limit reads,
\begin{eqnarray} 
\lim_{k\to 0}\bigg[ A h_0^1(kr)+B h_0^2(kr)\bigg]&&
\\\nonumber
&&\!\!\!\!\!\!\!\!\!\!\!\!\!\!\!\!\!\!\!\!\!=\lim_{k\to 0} \bigg[C(k) e^{-ikR}\left(\cosh(k'R)+\frac{i k}{k'}\sinh{k'R}\right)\frac{-ie^{ikr}}{kr}
\\\nonumber
&&+C(k) e^{ikR}\left(\cosh(k'R)-\frac{i k}{k'}\sinh{k'R}\right)\frac{ie^{ikr}}{kr}\bigg]~,
\\\nonumber
&&\!\!\!\!\!\!\!\!\!\!\!\!\!\!\!\!\!\!\!\!\!=\lim_{k\to 0} \bigg[\frac{-i(1-ikR)(\cosh(\alpha m R)+\frac{i k}{\alpha m}\sinh(\alpha m R))(1+ikr)}{2\sqrt{8 \pi^3}kr \cosh(\alpha m R)} 
\\\nonumber
&&+\frac{i(1+ikR)(\cosh(\alpha m R)-\frac{i k}{\alpha m}\sinh(\alpha m R))(1-ikr)}{2\sqrt{8 \pi^3}kr \cosh(\alpha m R)}\bigg]~,
\\\nonumber
&&\!\!\!\!\!\!\!\!\!\!\!\!\!\!\!\!\!\!\!\!\!=\frac{1}{\sqrt{8 \pi^3}}\left(1-\frac{\alpha m R-\tanh(\alpha m R)}{\alpha m r} \right)~.
\end{eqnarray}
In summary we have 
\begin{equation}
  \lim_{k\to 0}\Tilde{\phi}_{k}(\vec{r})=\frac{1}{\sqrt{8 \pi^3}} \phi_\mathrm{st}'(\vec{r})~,
\end{equation}
where $\phi_\mathrm{st}'(\vec{r})$ is the spatial part of the stationary solution given by Eq.~\eqref{phi-solution-quasistatic}.

For the second term in Eq.~\eqref{eq:preapprox}, we do an expansion around $k=0$ to first order. This approximation is justified, since the eigenfunctions with small $k$ are the ones that drive the evolution for long enough times, as we already discussed. The field is then given by

\begin{eqnarray}
\nonumber
\phi_0(\vec{r},t) \!\!&\approx& \!\!\phi_\mathrm{st}'(\vec{r})e^{-imt}+\frac{1}{2\pi^2}\int e^{-i(m+\frac{k^2}{2m})t}\phi_\mathrm{st}'(\vec{r}) \left(\frac{\alpha m R-\tanh(\alpha m R)}{\alpha m}\right)\,dk^3~,
\\
\phi_0(\vec{r},t) \!\!&\approx&\!\! \phi_\mathrm{st}'(\vec{r})e^{-imt}+\frac{2}{\pi}\phi_\mathrm{st}'(\vec{r})e^{-imt}\left(\frac{\alpha m R-\tanh(\alpha m R)}{\alpha m}\right)\int_0^{\infty} e^{-i\frac{k^2}{2m}t} \,dk~.
 \end{eqnarray}
 Doing a change of variable, $k^2=z$, and using that $\int_{-\infty}^{\infty} \frac{e^{-i z t}}{\sqrt{|z|}} \,dz=\sqrt{\frac{ 2 \pi}{t}}$, we arrive at
 \begin{equation}
\phi_0(\vec{r},t) \approx \phi_\mathrm{st}'(\vec{r})e^{-imt}+\phi_\mathrm{st}'(\vec{r})e^{-imt} \left(\frac{\alpha m R-\tanh(\alpha m R)}{\alpha} \right)\sqrt{\frac{2}{\pi m t}}~.
 \end{equation}
In this last expression, we just need to take the real part leading to obtain the final result
\begin{equation}
\phi_0(\vec{r},t) \approx \phi_\mathrm{st}(r,t)\left(1+ \frac{\alpha m R-\tanh(\alpha m R)}{\alpha} \sqrt{\frac{2}{\pi m t}}\right)~.
\label{sacadap}
\end{equation}
\section{The evolution of the field at the origin for a disappearing source}
\label{app:disappearing-source}

In this appendix we consider a situation where  where we have a spherical source present for times $t < 0$. At $t = 0$ we imagine that it disappears. Accordingly, the stationary solution, Eq.~\eqref{phi-solution-quasistatic} is the initial condition at $t=0$. At later times we expect that the solution converges to $\phi_\mathrm{homog}$.  
One benefit of considering this problem is that the differential equation for $t > 0$ is simply the free Klein-Gordon equation. This simplifies both an analytical as well as a numerical treatment.

\subsection{Analytical computation}
As mentioned, we take the field profile in Eq.~\eqref{phi-solution-quasistatic} to be the initial condition for the sourceless equation of motion. Now we make the ansatz
\begin{equation}
    \phi=\phi_{\infty} (\cos (m t+\delta) +\varphi(t,r))\;,
\end{equation}
and the sourceless equation of motion for $\varphi$ is 
\begin{equation}
    \ddot{\varphi}- \nabla^2 \varphi +m^2 \varphi =0~.
    \label{eq:KG}
\end{equation}
This is just a Klein-Gordon equation, so in principle we can solve the initial value problem with the (known) Green's function (following Morse and Feshbach~\cite{morsefeshbach}). So that 
\begin{equation}
    \varphi(r,t) = \frac{1}{4 \pi}\int dV\; \left[ G(r-r_0,t) \dot{\varphi}(t=0) -\left.\frac{\partial G}{\partial t_0}\right|_{t_0=0}\varphi(t=0)\right]~.
    \label{eq:soln}
\end{equation}
In practise however it is still hard to compute the general form of $\varphi(r,t)$ in this way. However what can be done is to compute the time evolution of the field at the origin, $r=0$.

We need the spatial Green's function for the Klein-Gordon equation
\begin{equation}
    G(r,t,r_0,t_0) = \frac{\delta(\tau - R)}{R}-\frac{m}{\sqrt{\tau^2-R^2}}J_1(m\sqrt{\tau^2-R^2})\Theta(\tau-R)\;,
\end{equation}
where $\tau =t-t_0$ and $R=|\vec{r}-\vec{r}_0|$, $J_1(x)$ is a Bessel function of the first kind and $\Theta(x)$ is the Heaviside step function.

As above, to compute the evolution of the field at $r=0$ we need the Green's function and its derivative with respect to $t_0$ when $t_0=0$ and $r=0$, these are
\begin{equation}
    G(t,r_0) = \frac{\delta(t -r_0)}{r_0}-\frac{m}{\sqrt{t^2-r_0^2}}J_1(m\sqrt{t^2-r_0^2})\Theta(t-r_0)\;,
\end{equation}
and
\begin{align}
    \frac{\partial G(t,r_0)}{\partial t_0} = & -\frac{\delta^{\prime}(t - r_0)}{r_0} -\frac{m t}{(t^2-r_0^2)^{3/2}}J_1(m\sqrt{t^2-r_0^2})\Theta(t-r_0)\nonumber\\
    & +\frac{m^2t}{2(t^2-r_0^2)}\left(J_0(m\sqrt{t^2-r_0^2})-J_2(m\sqrt{t^2-r_0^2})\right)\Theta(t-r_0)\nonumber\\
   & +\frac{m}{\sqrt{t^2-r_0^2}}J_1(m\sqrt{t^2 - r_0^2})\delta(t-r_0)\;.
\end{align}
Substituting these expressions into Equation (\ref{eq:soln}), the general solution of the initial value problem in terms of Green's functions, and choosing $r=0$,  is
\begin{align}
    \varphi(0,t) = & \varphi_i(t)+t\left.\frac{\partial \varphi_i}{\partial r}\right|_{r=t}+t \dot{\varphi}_i(t)\nonumber\\
    &+m\int dr_0\;r_0^2J_1(m\sqrt{t^2-r_0^2})\Theta(t-r_0)\left(-\frac{\dot{\varphi}_i(r_0)}{(t^2-r_0^2)^{1/2}}+\frac{t\varphi_i(r_0)}{(t^2-r_0^2)^{3/2}}\right)\nonumber\\
        &-m^2 \int dr_0\;r_0^2\frac{\varphi_i(r_0) t}{2(t^2-r_0^2)}\Theta(t-r_0)\left(J_0(m\sqrt{t^2-r_0^2})-J_2(m\sqrt{t^2-r_0^2})\right)\nonumber\\
        &-\frac{m^2t^2\varphi_i(t)}{2}\;.\label{eq:gfcalc}
\end{align}
If we take $m=0$ in Eq.~(\ref{eq:gfcalc}) we recover `Poisson's solution'. For $m\neq0$ it appears as if there are divergent integrals, but using relations between Bessel functions we can see that these cancel. We then find,
\begin{align}
    \varphi(0,t) = & \varphi_i(t)+t\left.\frac{\partial \varphi_i}{\partial r}\right|_{r=t}+t \dot{\varphi}_i(t)-\frac{m^2t^2\varphi_i(t)}{2}\nonumber\\
    &-m\int_0^t dr_0\;r_0^2\frac{\dot{\varphi}_i(r_0)}{(t^2-r_0^2)^{1/2}}J_1(m\sqrt{t^2-r_0^2})\nonumber\\
        &+m^2 \int_0^t dr_0\;r_0^2\frac{\varphi_i(r_0) t}{(t^2-r_0^2)}J_2(m\sqrt{t^2-r_0^2})\;.
\end{align}
If the spatial extent of the disappearing point source is small then we can approximate 
\begin{equation}
    \varphi_i=\frac{q_A R_A }{r}\cos (mt_0 +\delta)\;.
\end{equation}
Using this we can find the expression for $\varphi(0,t)$, by solving the integrals containing the Bessel functions with a change of variable $u^2 = t^2 -r_0^2$. After a small amount of algebra we obtain
\begin{equation}\label{eq:exactanalytical}
    \varphi(0,t)  =  - q_AR_Am \sin (m t_0+\delta)  J_0(mt) -q_AR_Am \cos (mt_0+\delta) J_1(mt)\;.
\end{equation}
At very early times this can be approximated by
\begin{equation}
    \varphi(0,t)=-q_AR_Am \sin (mt_0+ \delta)  +\mathcal{O}(q_AR_Am^2t)\;,
\end{equation}
and at late times 
\begin{align}
\label{eq:exactanalyticalapprox}
    \varphi(0,t)  =&  - q_AR_Am \sin (mt_0+ \delta)  \sqrt{\frac{2}{\pi m t }}\cos \left(mt -\frac{\pi}{4}\right)\nonumber\\
    &-q_AR_Am \cos (mt_0+\delta)  \sqrt{\frac{2}{\pi m t }}\cos \left(mt -\frac{3\pi}{4}\right)\;.
\end{align}
This exhibits the same $\sqrt{\frac{1}{t}}$-behavior that we also observe for the appearing source.

\subsection{Numerical treatment}
Solving the problem of the disappearing source follows essentially the same steps as in Sec.~\ref{subsec:numerical}.
\begin{figure}[t]
    \centering
    \includegraphics[width=0.45\textwidth]{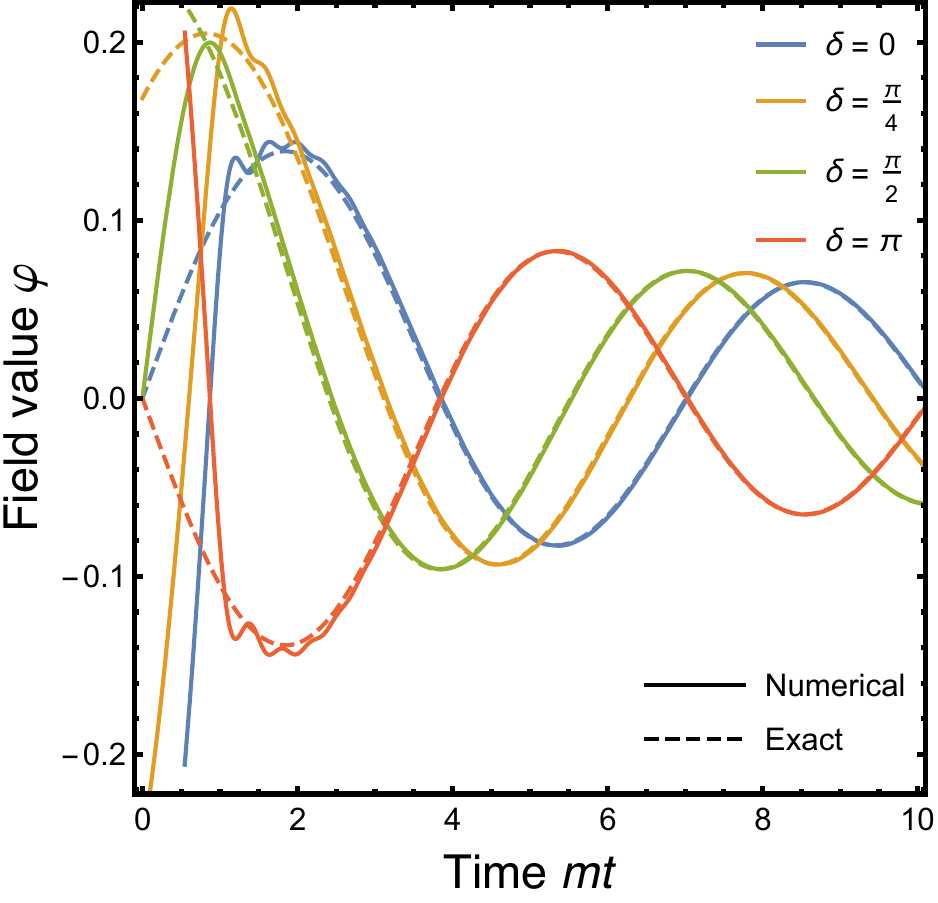}
    \hspace{1cm}
    \includegraphics[width=0.45\textwidth]{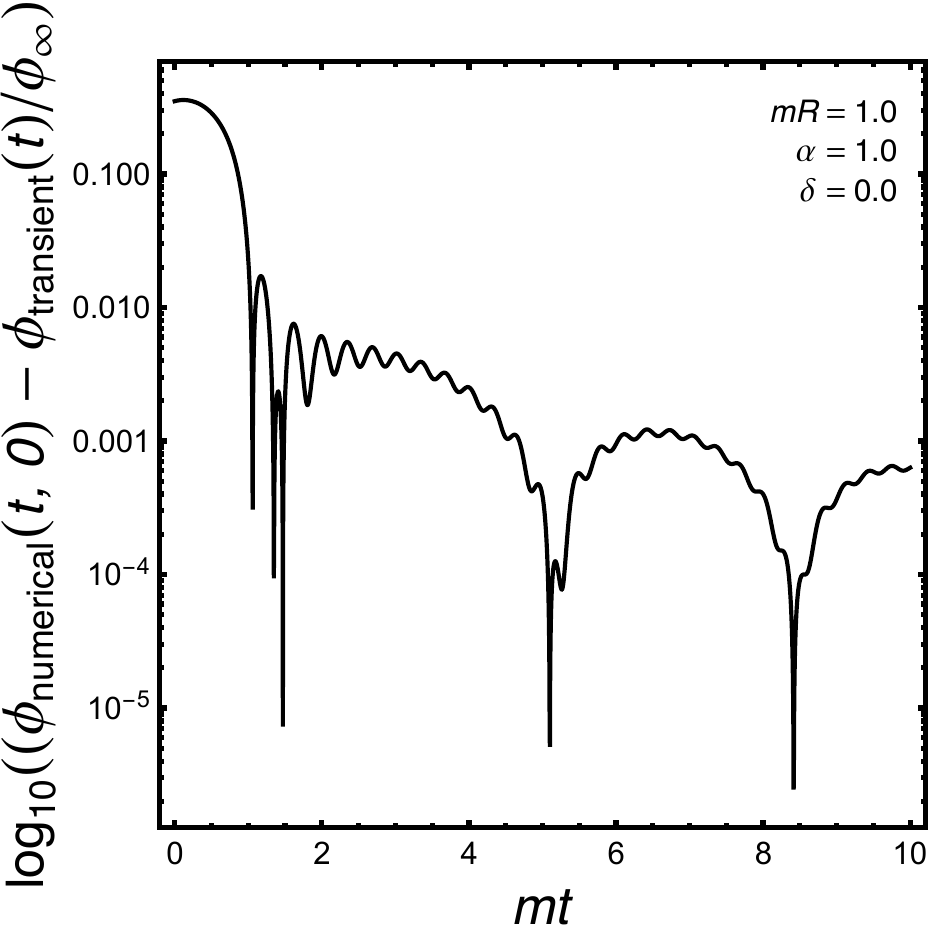}
    \caption{\small Numerical solutions of the central field value $\varphi(t, r = 0)$ for a disappearing source. {\bf Left:} Comparison between the numerical solutions and the analytical result Eq.~\eqref{eq:exactanalytical} for a range of phase offsets $\delta$.  The source has radius and density $m R = \frac{\rho}{m^2 M^2} = 1$. {\bf Right:} Difference between numerical solution and Eq.~\eqref{eq:exactanalytical} for a disappearing source.  This plot is characteristic of parameters $0.1 < m R < 10$ and $0.01 < \alpha < 10$, although the overall error increases with $\alpha$.}
    \label{fig:disappear-amplitude}
\end{figure}
Before solving the equation of motion, we perform a field transformation,
\begin{equation}
    \hat \phi = \hat \phi_\mathrm{homog} + \varphi~.
\end{equation}
Extracting the homogeneous solution makes it easier to assess the convergence at late times.  The deviation $\varphi$ is governed by the Klein-Gordon equation:
\begin{equation}
    \left(- \hat \Box - 1\right)\varphi = 0~.
\end{equation}
Note that the right hand side is zero only because $\alpha = 0$ for the times of interest $t > 0$.  This also implies that we are free to set the minimum spatial resolution $\hat{dr} = \hat R / 10$ irrespective of the source density.  {This may result in neglecting some initial very short-wavelength fluctuations near the origin for very dense sources, although the long-term behaviour should still be reliable, and we shall see shortly that this is indeed the case.}

The initial condition is then chosen to be
\begin{equation}
    \varphi(\hat t = 0, \hat r) = \hat \phi_\mathrm{st}(\hat t = 0, \hat r)  - \hat \phi_\mathrm{homog}(\hat t = 0)~,
\end{equation}
and likewise for the first derivative of $\varphi$.  The boundary conditions are similar to Eq.~\eqref{boundary-conditions-numerical} and read,
\begin{align} \nonumber
    \frac{d}{d\hat r} \varphi(\hat t, \hat r)\bigg |_{\hat r_\mathrm{max}} = 0~, \\
    \varphi(\hat t, \hat r_\mathrm{max}) = 0~.
    \label{BC-appearing-source}
\end{align}
We remind the reader that the boundary condition at $\hat r_\mathrm{max}$ is irrelevant as we will cut off the simulation before information of this point can propagate inwards.

\begin{figure}[!t]
    \centering
    \includegraphics[width=0.45\textwidth]{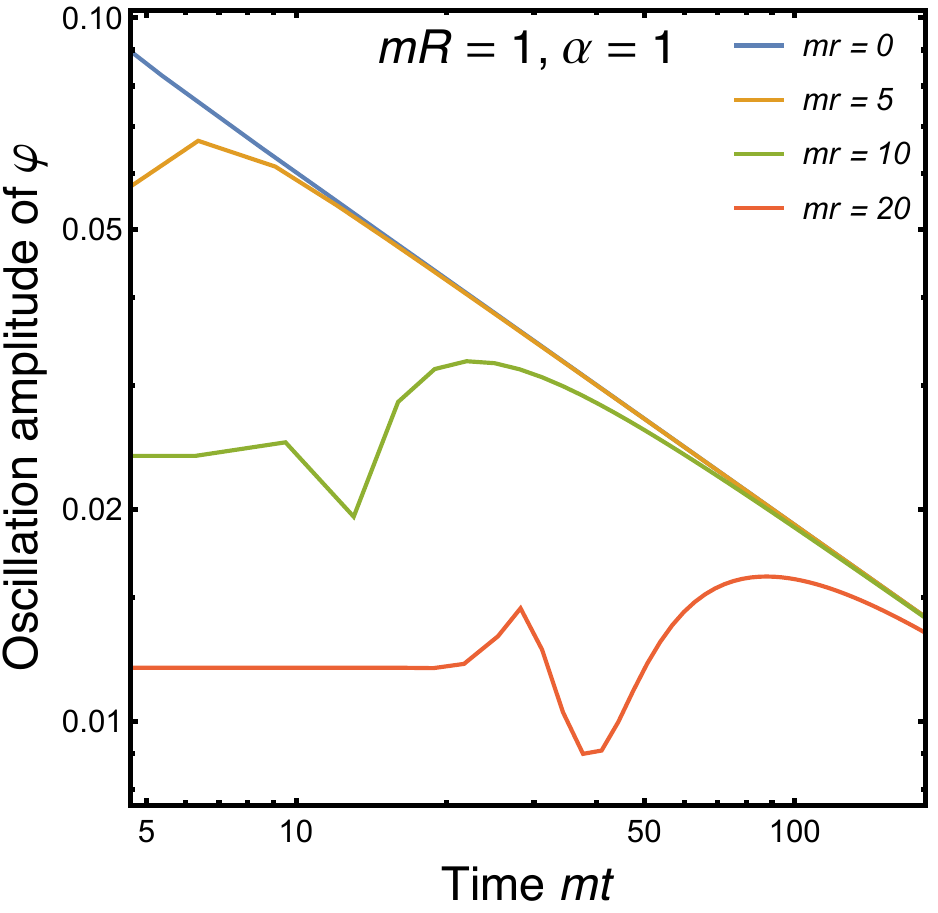}
    \hspace{1cm}
    \caption{\small Numerical results for the amplitude of the decaying sinusoid $\varphi(r, t)$ for a range of distances from the origin $r$.  We see that at late times all functions converge to the power law decay described at the origin by Eq.~\eqref{eq:exactanalyticalapprox}. Although we have shown the specific case $mR = \alpha = 1$ for concreteness, we have verified that this general pattern is characteristic of  all solutions with $\alpha, mR$ values in the range $[0.1, 10]$. }
    \label{fig:disappear-solutions}
\end{figure}

Sample numerical solutions for the central field value are shown in Fig.~\ref{fig:disappear-amplitude}.  We see excellent agreement between the numerical solutions and the exact transient solution given by Eq.~\eqref{eq:exactanalytical}.
Away from the origin, we no longer have an exact solution to compare to.  However, we might expect the decay pattern to be similar, which is that of a decaying sinusoid.  We plot the amplitude of this sinusoid vs. time for a range of radii in Fig.~\ref{fig:disappear-solutions}. We find that after an initial period, the amplitude of the sinusoidal function trends towards the same power-law decay, independent of radius.  Although we plot a single example, we have verified that the same qualitative behavior holds for $0.1 < \hat R < 10$, $0.1 < \alpha < 10$.
In other words, the solution at the origin, Eq.~\eqref{eq:exactanalyticalapprox}, provides a surprisingly accurate prediction of the late-time decay of the field, independent of $r$.

\bibliographystyle{utphys}
\bibliography{references}

\end{document}